\documentclass[draftclsnofoot,onecolumn]{IEEEtran}
\usepackage[english]{babel}
\usepackage{times,amssymb,amsmath,amsfonts,float,nicefrac,color,subfigure,algorithm2e}
\usepackage{euscript,graphics}
\usepackage[dvips]{epsfig}
\usepackage{bm}

\usepackage{multirow}

\ifCLASSINFOpdf
\else
\fi

\newtheorem{theorem}{Theorem}[section]
\newtheorem{lemma}[theorem]{Lemma}
\newtheorem{proposition}[theorem]{Proposition}

\newcommand{\remove}[1]{}

\newtheorem{cnstr}{Construction}

\newtheorem{xmpl}{Example}

\newcommand{\ff}{\mathbb{F}}

\newcommand\nc\newcommand
\nc\bfa{{\boldsymbol a}}\nc\bfA{{\bf A}}\nc\cA{{\mathcal A}}
\nc\bfb{{\boldsymbol b}}\nc\bfB{{\bf B}}\nc\cB{{\mathcal B}}
\nc\bfc{{\boldsymbol c}}\nc\bfC{{\bf C}}\nc\cC{{\mathcal C}}
\nc\bfd{{\boldsymbol d}}\nc\bfD{{\bf D}}\nc\cD{{\mathcal D}}
\nc\bfe{{\boldsymbol e}}\nc\bfE{{\bf E}}\nc\cE{{\mathcal E}}
\nc\bff{{\boldsymbol f}}\nc\bfF{{\bf F}}\nc\cF{{\mathcal F}}
\nc\bfg{{\boldsymbol g}}\nc\bfG{{\bf G}}\nc\cG{{\mathcal G}}
\nc\bfh{{\boldsymbol h}}\nc\bfH{{\bf H}}\nc\cH{{\mathcal H}}
\nc\bfi{{\boldsymbol i}}\nc\bfI{{\bf I}}\nc\cI{{\mathcal I}}
\nc\bfj{{\boldsymbol j}}\nc\bfJ{{\bf J}}\nc\cJ{{\mathcal J}}
\nc\bfk{{\boldsymbol k}}\nc\bfK{{\bf K}}\nc\cK{{\mathcal K}}
\nc\bfl{{\boldsymbol l}}\nc\bfL{{\bf L}}\nc\cL{{\mathcal L}}
\nc\bfm{{\boldsymbol m}}\nc\bfM{{\bf M}}\nc\cM{{\mathcal M}}
\nc\bfn{{\boldsymbol n}}\nc\bfN{{\bf N}}\nc\cN{{\mathcal N}}
\nc\bfo{{\boldsymbol o}}\nc\bfO{{\bf O}}\nc\cO{{\mathcal O}}
\nc\bfp{{\boldsymbol p}}\nc\bfP{{\bf P}}\nc\cP{{\mathcal P}}
\nc\bfq{{\boldsymbol q}}\nc\bfQ{{\bf Q}}\nc\cQ{{\mathcal Q}}
\nc\bfr{{\boldsymbol r}}\nc\bfR{{\bf R}}\nc\cR{{\mathcal R}}
\nc\bfs{{\boldsymbol s}}\nc\bfS{{\bf S}}\nc\cS{{\mathcal S}}
\nc\bft{{\boldsymbol t}}\nc\bfT{{\bf T}}\nc\cT{{\mathcal T}}
\nc\bfu{{\boldsymbol u}}\nc\bfU{{\bf U}}\nc\cU{{\mathcal U}}
\nc\bfv{{\boldsymbol v}}\nc\bfV{{\bf V}}\nc\cV{{\mathcal V}}
\nc\bfw{{\boldsymbol w}}\nc\bfW{{\bf W}}\nc\cW{{\mathcal W}}
\nc\bfx{{\boldsymbol x}}\nc\bfX{{\bf X}}\nc\cX{{\mathcal X}}
\nc\bfy{{\boldsymbol y}}\nc\bfY{{\bf Y}}\nc\cY{{\mathcal Y}}
\nc\bfz{{\boldsymbol z}}\nc\bfZ{{\bf Z}}\nc\cZ{{\mathcal Z}}

\DeclareSymbolFont{bbold}{U}{bbold}{m}{n}
\DeclareSymbolFontAlphabet{\mathbbold}{bbold}

\begin{document}
%
\title{Constructions of Optimal Cyclic $(r,\delta)$ Locally Repairable Codes}

\author{Bin~Chen,
        Shu-Tao~Xia,
        Jie~Hao, and Fang-Wei Fu  {\em Member, IEEE}
\thanks{The authors are with the Graduate School at Shenzhen, Tsinghua University, Shenzhen 518055, China. (emails: binchen14scnu@m.scnu.edu.cn, xiast@sz.tsinghua.edu.cn, j-hao13@mails.tsinghua.edu.cn, fwfu@nankai.edu.cn.)
This research is supported in part by the National Natural Science Foundation of China (61371078 and 61571243).}
\thanks{Bin Chen is with the School of Mathematical Sciences, South China Normal University, Guangzhou 510000, China. This research was done at the Graduate School at Shenzhen, Tsinghua University.}
\thanks{F.-W. Fu is with the Chern Institute of Mathematics and LPMC, Nankai University, Tianjin 300071, China .}}

\markboth{}%
{Shell \MakeLowercase{\textit{et al.}}: Bare Demo of IEEEtran.cls for IEEE Journals}

\maketitle
\begin{abstract}
A code is said to be a $r$-local locally repairable code (LRC) if each of its coordinates can be repaired by accessing at most $r$ other coordinates. When some of the $r$ coordinates are also erased, the $r$-local LRC can not accomplish the local repair, which leads to the concept of $(r,\delta)$-locality. A $q$-ary $[n, k]$ linear code $\cC$ is said to have $(r, \delta)$-locality ($\delta\ge 2$) if for each coordinate $i$, there exists a punctured subcode of $\cC$ with support containing $i$, whose length is at most $r + \delta - 1$, and whose minimum distance is at least $\delta$. The $(r, \delta)$-LRC can tolerate $\delta-1$ erasures in total, which degenerates to a $r$-local LRC when $\delta=2$. A $q$-ary $(r,\delta)$ LRC is called optimal if it meets the Singleton-like bound for $(r,\delta)$-LRCs. A class of optimal $q$-ary cyclic $r$-local LRCs with lengths $n\mid q-1$ were constructed by Tamo, Barg, Goparaju and Calderbank based on the $q$-ary Reed-Solomon codes. In this paper, we construct a class of optimal $q$-ary cyclic $(r,\delta)$-LRCs ($\delta\ge 2$) with length $n\mid q-1$, which generalizes the results of Tamo \emph{et al.} Moreover, we construct a new class of optimal $q$-ary cyclic $r$-local LRCs with lengths $n\mid q+1$ and a new class of optimal $q$-ary cyclic $(r,\delta)$-LRCs ($\delta\ge 2$) with lengths $n\mid q+1$.
The constructed optimal LRCs with length $n=q+1$ have the best-known length $q+1$ for the given finite field with size $q$ when the minimum distance is larger than $4$.
\end{abstract}


%
\IEEEpeerreviewmaketitle

\section{Introduction}
In distributed storage systems, repair cost metrics include repair locality \cite{gopalan2011locality,locality2}, repair bandwidth \cite{network} and disk-I/O \cite{MDS array codes}. Recently, locally repairable codes (LRCs) introduced by Gopalan \emph{et al.} \cite{gopalan2011locality} have attracted a lot of interest. The $i${th} symbol $c_i$ of an $[n, k]$ linear code $\cC$ over the finite field $\ff_{q}$  is said to have \emph{locality} $r$ if this symbol can be recovered by accessing at most $r$ other symbols of $\cC$. Coding techniques are then introduced in distributed storage systems to reduce the storage overhead, while maintaining high data reliability. Maximum distance separable (MDS) codes can be used as erasure codes in distributed storage systems and any symbol can be recovered by accessing any $k$ other symbols. In order to reduce the repair costs in distributed storage systems, the locality parameter $r \ll k$ is often demanded, which implies that only a small number of storage nodes are involved in repairing a failed node. The code is called a $q$-ary $(n,k,r)$ LRC with all symbol locality $r$ or a $r$-local LRC for brevity if all the $n$ symbols have locality $r$. The Singleton-like bound of the minimum distance $d$ for an $(n, k, r)$ LRC \cite{gopalan2011locality} said that
\begin{equation}
\label{singleton}
d\le n-k-\left\lceil\frac{k}{r}\right\rceil+2,
\end{equation}
where $\lceil \cdot \rceil$ denotes the ceiling function. The codes meeting the above bound (\ref{singleton}) are called optimal $r$-local LRCs.
Various constructions of optimal $r$-local LRCs  were obtained recently, e.g., \cite{gopalan2011locality}, \cite{Tamo 2013ISIT}-\cite{Zeh ITW}, \cite{some-results}. To the  best of our knowledge, for a given finite field with size $q$, the code length $n$ was not larger than $q$ in all the known constructions except the ones in \cite{some-results} where the minimum distance $d=2$ or $d=4$.

Cyclic LRCs were studied very recently. Goparaju and Calderbank \cite{gop14} constructed new families of binary cyclic codes that have an optimal dimension for given minimum distance $d$ and locality $r$, including $r=2$ and $d=2, 6, 10$. Huang \emph{et al.} \cite{PHuang ITW} analyzed the locality of many traditional cyclic codes, e.g., Hamming code, Simplex codes, and BCH codes. Constructions of optimal cyclic codes in terms of the dimension for given distance and length over small field were discussed in \cite{Zeh ITW}. Tamo, Barg, Goparaju and Calderbank \cite{Tamo cyclic} focused on the cyclic LRCs in terms of their zeros. A class of optimal $q$-ary cyclic LRCs with length $n\mid q-1$ were then constructed by analyzing the structure of zeros of Reed-Solomon codes and cyclic LRCs. They also studied the upper bound of the locality parameter $r$ for the subfield subcodes of cyclic LRCs, which was equivalent to estimate the dual distance $d^{\bot}$.

When some of the $r$ repairing symbols are also erased, the $r$-local LRC can not accomplish the local repair, which leads to the concept of $(r,\delta)$-locality.
Prakash \emph{et al.} \cite{prakash2012optimal} addressed the situation of multiple device failures and gave a kind of generalization of $r$-local LRCs. According to \cite{prakash2012optimal}, the $i${th} symbol $c_i$ of a $q$-ary $[n, k]$ linear code $\cC$ is said to have $(r, \delta)$-locality ($\delta\ge2$) if  there exists a punctured subcode of $\cC$ with support containing $i$, whose length is at most $r + \delta - 1$, and whose minimum distance is at least $\delta$, i.e., there exists a subset $S_{i}\subseteq [n]\triangleq\{1,2,\ldots,n\}$ such that $i\in S_{i}$, $|S_{i}|\le r+\delta-1$ and $d_{min}(\cC|_{S_{i}})\ge\delta$. The code $\cC$ is said to have $(r, \delta)$  locality or be a $(r,\delta)$-LRC if all the symbols have $(r,\delta)$ localities. A generalized Singleton-like bound was also obtained in \cite{prakash2012optimal}, which said that the minimum distance $d$ of a $(r, \delta)$-LRC is upper bounded by
\begin{equation} \label{eq_GeneralizedSingleton}
d \leq n-k+1-\left( \left \lceil \frac{k}{r} \right \rceil-1 \right)(\delta -1).
\end{equation}
The codes meeting the above bound (\ref{eq_GeneralizedSingleton}) are called optimal $(r,\delta)$-LRCs.
Note that when $\delta=2$, the notion of locality in \cite{prakash2012optimal} reduces to the notion of locality in \cite{gopalan2011locality}.
In \cite{prakash2012optimal}, a class of optimal $(r, \delta)$-LRCs with length $n=\lceil\frac{k}{r}\rceil(r+\delta-1)$ were obtained for $q>n$, and there exist optimal $(r, \delta)$-LRCs when $r+\delta-1\mid n$ and $q > kn^{k}$. An algebraic construction of optimal $(r, \delta)$-LRCs with $q\geqslant  n$ was proposed in \cite{tam14a} based on polynomial evaluations. By studying the structure of matroid represented by the optimal LRC's generator matrix, optimal $(r, \delta)$-LRCs were obtained in \cite{Tamo matroidIT} with $q\geqslant (\frac{nr}{r+\delta-1})^{ak+1}$. The existence conditions and deterministic construction algorithms for optimal $(r, \delta)$-LRCs with $q\ge \binom{n}{k-1}$ were given in \cite{song14}. Based on a known optimal $(r, \delta)$-LRC, \cite{Ernvall2016IT} obtained more optimal $(r, \delta)$-LRCs by lengthening or shortening. To the best of our knowledge, the existing results on cyclic $(r, \delta)$ LRCs are limited to the special case of $\delta=2$ or the cyclic $r$-local LRCs stated in the last paragraph.
There are also other generalizations of $r$-local LRCs, e.g., the vector codes with locality \cite{papil}-\cite{Ernvall2016IT}, and the $t$-available-$r$-local or $(r,t)$ LRCs \cite{Wang}-\cite{WangCon}. However, this paper will be limited to the $(r,\delta)$-LRCs.

There has been a famous problem for a long time related to the MDS conjecture \cite{MacWilliams}: for the given finite field size $q$ and dimension $k$, find the largest value of $n$ for which there exists a non-trivial $q$-ary MDS code with length $n$. Although there is no answer to it up to now, the evidence seems to suggest that the largest value of $n$ is actually $q+1$ \cite{Roman}. Cyclic MDS codes, especially the cyclic Reed-Solomon Codes with length $q-1$ and the Berlekamp-Justesen codes with length $q+1$ \cite{Berlekamp,MacWilliams}, are among the most important MDS codes. The similar situation seems to lie in the area of optimal LRCs. As stated above, Tamo, Barg, Goparaju and Calderbank \cite{Tamo cyclic} constructed a class of $q$-ary cyclic $r$-local LRCs with length $n\mid q-1$ based on the cyclic Reed-Solomon Codes.  In this paper, this elegant result is firstly generalized to the cases of $(r,\delta)$ LRCs. In fact, we obtain a class of optimal cyclic $(r,\delta)$-LRCs ($\delta\ge 2$) with length $n\mid q-1$. Moreover, we obtain a class of new optimal cyclic $(r,\delta)$-LRCs ($\delta\ge 2$) with longer length $n\mid q+1$ based on the Berlekamp-Justesen codes, while the case of $\delta=2$ indicates the $r$-local LRCs with length $n\mid q+1$. Comparing with the corresponding MDS codes, it seems difficult to obtain optimal cyclic LRCs with length larger than $q+1$. To the best of our knowledge, in all the known optimal constructions of LRCs, the length $n$ is not greater than the given size $q$ of finite field $\ff_{q}$  when the minimum distance is larger than $4$. Therefore, the proposed optimal constructions address the problem of constructing longer optimal LRCs restricted by the size of the alphabet, and could provide optimal LRCs with length $n=q+1$ greater than $q$ for any given finite field with size $q$.

The rest of paper is organized as follows. Section II gives some preliminaries on cyclic codes, MDS codes and some results of cyclic $r$-local LRCs in \cite{Tamo cyclic}. In Section \ref{q-1}, by generalizing the construction of optimal cyclic $r$-local LRCs in \cite{Tamo cyclic}, we construct a class of optimal cyclic $(r, \delta)$-LRCs with length $q-1$ and its factors. In Section \ref{q+1}, we firstly give constructions of optimal cyclic $r$-local LRCs with length $n\mid q+1$ for even and odd $q$, respectively, then further generalize them to construct a class of optimal cyclic $(r, \delta)$-LRCs with length $n\mid q+1$. Finally, Section \ref{conclusion} concludes the paper.

\section{Preliminaries}
\subsection{Cyclic codes and the BCH bound}
Let $\ff_{q}$ be a finite field with size $q$, where $q$ is a prime power. A cyclic code is an ideal of the ring $\ff_q[x]/(x^n-1)$, where $(n, q)=1$ \cite{MacWilliams,Roman}. Let $s$ be the order of $q$ modulo $n$, that is, the least number of $i$ such that $n\mid (q^i-1)$. Let $F_{q^s}$ be the splitting field of $x^n-1$. Let $\beta$ be a primitive element of $F_{q^s}$ and $\alpha=\beta^{(q^s-1)/n}$ be a primitive $n$-th root of unity. Let $\cC$ be a $q$-ary  $[n, k, d]$  cyclic code with generator polynomial $g(x)$, where $g(x)\mid x^n-1$. It is well known that every cyclic shift of any codeword of $\cC$ is still in $\cC$ and $\deg g(x)=n-k$. The zeros set $Z=\{ \alpha^{i_{j}} \mid g(\alpha^{i_{j}})=0, \;j=1, 2, \dots, n-k\}$ of $g(x)$ is called the \emph{complete defining set} of $\cC$.
The next result is the well-known generalized BCH bound of cyclic codes \cite{Roman}.
\begin{lemma}
\label{bchbound}
Let $\cC$ be a $q$-ary cyclic code with generator polynomial $g(x)$, and $\alpha$ be a primitive $n$-th root of unity. If $g(x)$ has
$$\alpha^{u}, \alpha^{u+b}, \dots, \alpha^{u+(\delta-2)b}$$
among its zeros, where $b$ and $n$ are relatively prime and $u\ge0$. Then  the minimum distance of $\cC$ is at least $\delta$.
\end{lemma}

\subsection{Cyclic MDS codes }
The well known examples of cyclic MDS codes over $\ff_{q}$ are the Reed-Solomon codes of length $q-1$ or its factors. Based on Reed-Solomon codes, Tamo \emph{et al.} \cite{Tamo cyclic} constructed optimal cyclic $r$-local LRCs with length $n\mid q-1$.
Roth and Seroussi \cite{Roth} showed that  nontrivial cyclic MDS codes of length $q$ over $\ff_{q}$ exist if and only if $q$ is prime. Berlekamp and Justesen \cite{Berlekamp} introduced a class  of  cyclic MDS codes of length $q+1$ ($q=2^m$), which can also be found in \cite{MacWilliams}. But there is a small mistake in \cite{MacWilliams} which says that a similar construction exists for odd $q$ and arbitrary $k$.  Actually, due to the results in \cite{Georgiades,Cyclic and Pseudo}, when $q$ is odd and $n$ is a factor of $q+1$, nontrivial cyclic $[n, k]$ MDS codes over $\ff_{q}$ do not exist if both $n$ and $k$ are even. These cyclic MDS codes with length $q+1$ were the known nontrivial constructions with largest length for a given $q$. Although cyclic MDS codes with length $n\mid q+1$ over $\ff_{q}$ can be regarded as  subfield subcodes of a Reed-Solomon code with length $n\mid q^{2}-1$ over $\ff_{q^{2}}$, the structure of zeros of the cyclic MDS code with length $n\mid q+1$ over $\ff_{q}$ is of much interest and should be carefully divided into several cases, which is listed in detail as follows.

Let $\ff_{q^2}$ be the extension field of $\ff_{q}$ and $\beta$ be a primitive element of $\ff_{q^2}$. Let $n\mid q+1$ and $\alpha=\beta^{(q^2-1)/n}$ be a primitive $n$-th root of unity. Then for any positive integer $i$, it is easy to see that $\alpha^{-i}+\alpha^{i}\in \ff_{q}$. Note that $\alpha^{-i}=\alpha^{n-i}=\alpha^{qi}$, and for a $q$-ary cyclic code with length $n$, $\alpha^i$ is its zero if and only if $\alpha^{-i}$ is also its zero.
When $q$ is even, $n$ has to be odd and the roots of $x^{n}-1$ can be formulated as $\alpha^{-{(n-1)}/{2}}$, $\alpha^{-{(n-3)}/{2}}$, $\dots, \alpha^{-1}, \alpha^{0}, \alpha^{1},\dots, \alpha^{{(n-3)}/{2}}, \alpha^{{(n-1)}/{2}} $. Furthermore,
$$x^{n}-1=(x-1)f_{1}(x)f_{2}(x)\dots f_{{(n-1)}/{2}}(x),$$
where $f_{i}(x)=x^{2}-(\alpha^{-i}+\alpha^{i})x+1=(x-\alpha^{-i})(x-\alpha^{i})$ are irreducible quadratics over $\ff_{q}$. For any $1 < k < n$, the consecutive zeros of $g(x)$ can be seen in Table I to obtain  $[n, k, d]$ cyclic MDS codes. When $q$ is odd and $n$ is odd, for any $1 < k < n$, the consecutive zeros of $g(x)$ can be taken as the previous case. When $q$ is odd, $n$ is even, and $k$ is odd, the zeros of $x^{n}-1$ can be formulated as $\alpha^{-n/2+1},\dots, \alpha^{-1}, \alpha^{0}, \alpha^{1},\dots, \alpha^{n/2-1}, \alpha^{n/2} $. Furthermore,
$$x^{n}-1=(x-1)(x-\alpha^{{n}/{2}})f_{1}(x)f_{2}(x)\dots f_{({n}/{2}-1)}(x).$$
The consecutive zeros of $g(x)$ can also be seen in Table I to obtain  $[n, k]$ cyclic MDS codes. These structures of zeros of cyclic MDS Codes are crucial for the constructions of optimal $q$-ary cyclic LRCs in the following sections.

\begin{table}[!hbp]\label{table1}
\centering
\caption{\bf Cyclic MDS Codes over $\ff_{q}$ with length $n\mid q+1$ }
\begin{tabular}{|c|c|}
\hline
 Conditions & Complete defining set  \\
\hline
$q$ even, $k$ even; &\multirow{2}{*}{$\lbrace \alpha^{i} \mid -\frac{n-1-k}{2}\le i\le \frac{n-1-k}{2}\rbrace$} \\
 or $q$ odd, $k$ even, $n$ odd&  \\
\hline
$q$ even, $k$ odd; &\multirow{2}{*}{$\lbrace \alpha^{i}\mid \frac{k+1}{2}\le i\le \frac{2n-1-k}{2}\rbrace$}\\
 or $q$ odd, $k$ odd, $n$ odd&\\
 \hline
\multirow{4}{*}{$q$ odd, $k$ odd, $n$ even}&\multirow{2}{*}{$\lbrace  \alpha^{i}\mid -\frac{n-1-k}{2}\le i\le \frac{n-1-k}{2}\rbrace$}\\
&  \\
 \cline{2-2}
&\multirow{2}{*}{$\lbrace \alpha^{i}\mid \frac{k+1}{2}\le i\le \frac{2n-1-k}{2}\rbrace$}\\
&  \\
 \hline
\end{tabular}
\end{table}

\subsection{Optimal cyclic  $r$-local LRCs with length $q-1$ and its factors}
Tamo, Barg, Goparaju and Calderbank \cite{Tamo cyclic} set up an elegant characterization framework and gave detailed constructions for cyclic $r$-local LRCs, some of which, e.g., Lemma 3.3, Proposition 3.4, and Theorem 3.1 in \cite{Tamo cyclic}, are recalled here and will be employed in the later sections.

\begin{lemma}\label{lem 2.2}
Let $s$ be the order of $q$ modulo $n$ and $\ff_{q^s}$ be the splitting field of $x^n-1$. Let $\alpha\in \ff_{q^s}$ be a primitive $n$-th root of unity. Let $r$ be a positive integer such that $(r+1)\mid n$ and $l$, $0\le l\le r$ be an integer.
Consider a $\nu\times n$ matrix $\cH$ with the rows
  $$
  h_{m}=(1, \alpha^{m(r+1)+l},\alpha^{2(m(r+1)+l)},\dots,\alpha^{(n-1)(m(r+1)+l)}),
  $$
where $m=0,1,\dots,\nu-1$, and $\nu=n/(r+1)$.
Then all the cyclic shifts of the $n$-dimensional vector of weight $r+1$
   $$
v=(1\underbrace{0 \ldots 0}_{\nu-1}\alpha^{l\nu}\underbrace{0\ldots0}_{\nu-1}\alpha^{2l\nu}\underbrace{0\ldots0}_{\nu-1}\ldots
\alpha^{rl\nu}\underbrace{0\ldots0}_{\nu-1})
  $$
are contained in the row space of $\cH$  over $\ff_{q^s}$.
\end{lemma}
\begin{lemma} \label{pro 2.3}
Let $\cC$ be a cyclic code of length $n$ over $\ff_q$ with the complete defining set $Z$, and
let $r$ be a positive integer such that $(r+1)\mid n.$
If $Z$ contains some coset of the group of $\nu$-th roots of unity, where $\nu=n/(r+1)$.
Then $\cC$ has locality  at most $r$.
\end{lemma}
\begin{lemma}\label{thm:cyclic}
Let $\alpha\in \ff_q$ be a primitive $n$-th root of unity, where $n\mid (q-1)$. Assume that $r\mid k$ and $\mu=k/r$. Let $l, 0\le l\le r$ be an integer and $b\ge 1$ be an integer such that $(b,n)=1$.
Consider the following sets of elements of $\ff_q$:

  $
  L=\{\alpha^i, i\,\text{mod}(r+1)=l\},
  $
  and

  $
  D=\{\alpha^{j+sb}, s=0,\dots,n-{\mu}({r+1})\},
  $
  \\
  where $\alpha^j\in L.$
The cyclic code with the defining set of zeros $L\cup D$ is an optimal $(n,k,r)$ $q$-ary cyclic LRC code.
\end{lemma}

In order to construct cyclic $r$-local LRCs, \cite{Tamo cyclic} divided the complete defining set into $L\cup D$, where $L$ is the locality zero set to ensure locality and $D$ is the consecutive zero set to ensure large minimum distance. We call this the \emph{$L\cup D$ construction} in the rest of the paper.

\section{Optimal cyclic  $(r, \delta)$-LRCs with length $q-1$ and its factors}\label{q-1}

In this section, we will generalize the construction of optimal cyclic $r$-local LRCs in \cite{Tamo cyclic} to the case of $(r, \delta)$-LRCs ($\delta\ge 2$), which might be the first class of optimal cyclic $(r,\delta)$-LRCs to the best of our knowledge.
Throughout this section, assume that $\delta\ge 2$, $n\mid(q-1)$, $r+\delta-1\mid n$. Let $\alpha\in\ff_q$ be a primitive $n$-th root of unity and let $$
    \nu^{\prime}=n/(r+\delta-1), \quad \mu=k/r \;\mbox{ if }\; r\mid k.
  $$

Similar to Lemma \ref{lem 2.2},  the following lemma is obvious.
\begin{lemma}\label{lemma 4.1} Let $0\le i_{1}<i_{2}<\dots <i_{\delta-1}\le r+\delta-2$ be an arithmetic progression with $\delta-1$ items and common difference $b$, where $(b,n)=1$. Consider a $(\delta-1)\nu^{\prime}\times n$ matrix $\cH$ with the rows
\begin{eqnarray*}
  h_{m}^{(j)}&=&(1, \alpha^{m(r+\delta-1)+i_{j}},\alpha^{2(m(r+\delta-1)+i_{j})}, \dots,\alpha^{(n-1)(m(r+\delta-1)+i_{j})}),
\end{eqnarray*}
 where $j=1, 2, \dots, \delta-1$, $m=0,1,\dots,\nu^{\prime}-1,$ and $\nu^{\prime}=n/(r+\delta-1).$ Then  all the cyclic shifts of the row vectors of weight $r+\delta-1$ in the following  $(\delta-1)\times n$-matrix $V$:
   $$
 \begin{pmatrix}
1\underbrace{0 \ldots 0}_{\nu^{\prime}-1}&\alpha^{{\nu^{\prime} i_{1}}}&\underbrace{0\ldots0}_{\nu-1}&(\alpha^{\nu^{\prime} i_{1}})^{2}&\ldots
&(\alpha^{\nu^{\prime} i_{1}})^{r+\delta-2}&\underbrace{0\ldots0}_{\nu^{\prime}-1}\\
1\underbrace{0 \ldots 0}_{\nu^{\prime}-1}&\alpha^{\nu^{\prime} i_{2}}&\underbrace{0\ldots0}_{\nu^{\prime}-1}&(\alpha^{\nu^{\prime} i_{2}})^{2}&\ldots
&(\alpha^{\nu^{\prime} i_{2}})^{r+\delta-2}&\underbrace{0\ldots0}_{\nu^{\prime}-1}\\
\vdots&\vdots&\vdots&\vdots&\vdots&\vdots&\vdots\\
1\underbrace{0 \ldots 0}_{\nu^{\prime}-1}&\alpha^{\nu^{\prime} i_{\delta-1}}&\underbrace{0\ldots0}_{\nu^{\prime}-1}&(\alpha^{\nu^{\prime} i_{\delta-1}})^{2}&\ldots
&(\alpha^{\nu^{\prime} i_{\delta-1}})^{r+\delta-2}&\underbrace{0\ldots0}_{\nu^{\prime}-1}\\
\end{pmatrix}
  $$
are contained in the row space of $\cH$ over $\ff_q$.
\end{lemma}
\begin{IEEEproof}
Note that when $n\mid q-1$, the order of $q$ modulo $n$ is $1$, which implies that the the splitting field of $x^{n} -1$ is $\ff_q$. Similar to Lemma \ref{lem 2.2}, it is  easy to obtain that for any fixed $j\in\{1, 2, \dots, \delta-1\}$,  the $n$-dimensional vector of weight $r+\delta-1$
   $$
v_{j}=(1\underbrace{0 \ldots 0}_{\nu^{\prime}-1}\alpha^{\nu^{\prime} i_{j}}\underbrace{0\ldots0}_{\nu^{\prime}-1}(\alpha^{\nu^{\prime} i_{j}})^{2}\underbrace{0\ldots0}_{\nu^{\prime}-1}\ldots
(\alpha^{\nu^{\prime} i_{j}})^{r+\delta-2}\underbrace{0\ldots0}_{\nu^{\prime}-1})
  $$
are contained in the row space of $\cH$  over $\ff_{q}$.
Then all the cyclic shifts of the row vectors of weight $r+\delta-1$ in the  $(\delta-1)\times n$-matrix $V$ are contained in the row space of $\cH$ over $\ff_q$.
\end{IEEEproof}

If we take the submatrix $V^{\prime}$ of $V$ formed by the non-zero columns:
  $$
 \begin{pmatrix}
1&\alpha^{\nu^{\prime} i_{1}}&(\alpha^{\nu^{\prime} i_{1}})^{2}&\ldots
&(\alpha^{\nu^{\prime} i_{1}})^{r+\delta-2}\\
1&\alpha^{\nu^{\prime} i_{2}}&(\alpha^{\nu^{\prime} i_{2}})^{2}&\ldots
&(\alpha^{\nu^{\prime} i_{2}})^{r+\delta-2}\\
\vdots&\vdots&\vdots&\vdots&\vdots\\
1&\alpha^{\nu^{\prime} i_{\delta-1}}&(\alpha^{\nu^{\prime} i_{\delta-1}})^{2}&\ldots
&(\alpha^{\nu^{\prime} i_{\delta-1}})^{r+\delta-2}\\
\end{pmatrix}
  $$
Then $V^{\prime}$ forms a parity-check matrix of a $[r+\delta-1, r, \delta]$ Reed-Solomon code. And the cyclic shifts of the row vectors in $V$ partition the support of the code into disjoint subsets of size $r+\delta-1$, which define the $(r, \delta)$-local recovering
sets of the symbols and satisfy the structure theorem in \cite{prakash2012optimal,song14}. Therefore, we obtain the following proposition similar to Lemma \ref{pro 2.3}.
\begin{proposition} \label{prop:cor1}
Let $\cC$ be a cyclic code of length $n$ over $\ff_q$ with the complete defining set $Z$, and
let $r$, $\delta$ be positive integers such that $(r+\delta-1)\mid n$. Let $0\le i_{1}<i_{2}<\dots <i_{\delta-1}\le r+\delta-2$ be an arithmetic progression with $\delta-1$ items and common difference $b$, where $(b,n)=1$.
If $Z$ contains some cosets of the group of $\nu^{\prime}$-th roots of unity $\cup_{m=i_{1}}^{i_{\delta-1}}L_{m}$,
where  $
  L_{m}=\{\alpha^i\mid i \mod (r+\delta-1)=m\}
  $,
Then $\cC$ has $(r, \delta)$-locality.
\end{proposition}
\begin{IEEEproof}
If $Z$ contains some cosets of the group of $\nu^{\prime}$-th roots of unity $\cup_{m=i_{1}}^{i_{\delta-1}}L_{m}$,  the row vectors:
 \begin{eqnarray*}
  h_{m^{\prime}}^{(j)}&=&(1, \alpha^{m^{\prime}(r+\delta-1)+i_{j}},\alpha^{2(m^{\prime}(r+\delta-1)+i_{j})}, \dots,\alpha^{(n-1)(m^{\prime}(r+\delta-1)+i_{j})}),
\end{eqnarray*}
 where $j=1, 2, \dots, \delta-1$, $m^{\prime}=0,1,\dots,\nu^{\prime}-1,$ and $\nu^{\prime}=n/(r+\delta-1)$,  are
contained in the row space of  the parity check matrix $\cH$ of $\cC$. Then by Lemma \ref{lemma 4.1} and the previous description, $\cC$ has $(r, \delta)$-locality.
\end{IEEEproof}

\begin{cnstr}\label{construction 1}
Let $\alpha \in \ff_q$ be a primitive $n$-th root of unity, where $n\mid(q-1)$. Let $0\le i_{1}<i_{2}<\dots <i_{\delta-1}\le r+\delta-2$ be an arithmetic progression with $\delta-1$ items and common difference $b$, where $(b,n)=1$. Suppose $r\mid k$ and let $\mu=k/r.$
Consider the following sets of elements of $\ff_q$:

  $
  L_{m}=\{\alpha^i\mid i \mod (r+\delta-1)=m\}, m=i_{1}, i_{2}, \dots, i_{\delta-1}
  $
  and

  $
  D=\{\alpha^{j+sb}\mid  s=0, 1, \dots,n-{\mu}({r+\delta-1})+\delta-2\},
  $
  \\
  where $\alpha^j\in L_{i_1}.$
Then the cyclic code $\cC$ with the complete defining set of zeros $(\cup_{m=i_{1}}^{i_{\delta-1}}L_{m})\cup D$ is a $q$-ary optimal cyclic $(r, \delta)$-LRC with length $n$, dimension $k$, and minimum distance $d=n-k+1-(\mu-1)(\delta-1)$.
\end{cnstr}
\begin{IEEEproof}
By Proposition \ref{prop:cor1}, $\cC$ has $(r, \delta)$-locality. Next, we calculate the dimension of $\cC$. Since $(b, n)=1$ implies $(b, r+\delta-1)=1$, for $m=i_{1}, i_{2}, \dots, i_{\delta-1}$,
\begin{eqnarray*}
|L_{m}\cap D|
&=&\left\lfloor\frac{n-\mu(r+\delta-1)+\delta-2}{r+\delta-1}\right\rfloor+1\;=\; \nu^{\prime}-\mu+1.
\end{eqnarray*}
Thus, after a few simple calculations, we have
\begin{eqnarray*}
\quad&&|(\cup_{m=i_{1}}^{i_{\delta-1}}L_{m})\cup D|\\
&=&|\cup_{m=i_{1}}^{i_{\delta-1}}L_{m}|+|D|-|(\cup_{m=i_{1}}^{i_{\delta-1}}L_{m})\cap D|\\
&=&(\delta-1)|L_{m}|+|D|-(\delta-1)|L_{m}\cap D|\\
&=&n-k.
\end{eqnarray*}
Hence, the generator polynomial has degree $n-k$, which implies that the dimension of $\cC$ is $k$.
Finally, the minimum distance $d$ of  $\cC$ is obtained by the generalized BCH bound in Lemma \ref{bchbound} for the set of consecutive zeros $D$ and the Singleton-like bound (\ref{eq_GeneralizedSingleton}) for $(r,\delta)$-LRCs.
\end{IEEEproof}

{\em Remark 1:}
It is not difficult to verify that the condition $r\mid k$ can be removed from Construction \ref{construction 1} by letting $D$ contain $n-k-(\lceil k/r\rceil-1)(\delta-1)$ zeros. We call the above construction \emph{$(\cup_{m}L_{m})\cup D$ construction} in the rest of the paper.
\begin{center}
\includegraphics[width=4.0in]{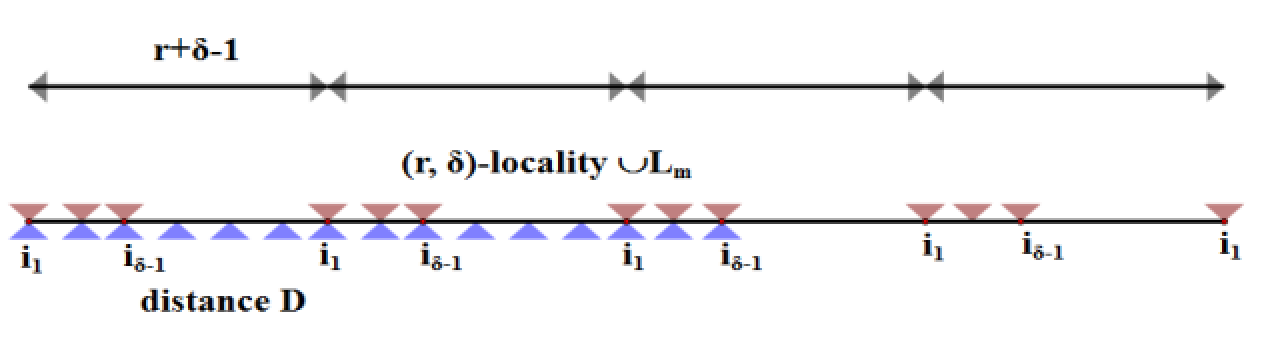}

{\small Fig.~1: Subsets of zeros for $D$  and $(r, \delta)$-locality for $\cup_{m=i_{1}}^{i_{\delta-1}}L_{m}$ $(b=1)$}
\end{center}
\begin{center}
\includegraphics[width=4.0in]{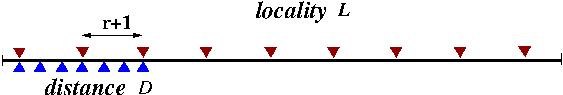}

{\small Fig.~2: Subsets of zeros for distance $D$ and $r$-local locality $L$ (Shown in \cite{Tamo cyclic} for $b=1$)}
\end{center}

{\em Remark 2:}
 When $\delta=2$, Construction \ref{construction 1} reduces to Theorem 3.1 in \cite{Tamo cyclic}. It is easy to see that the point $i_1$ will coincide with the point $i_{\delta-1}$ in Fig.~1$\thicksim$Fig.~2. So  Construction \ref{construction 1} is indeed a generalized construction of that in Theorem 3.1 in \cite{Tamo cyclic}.

\section{Optimal Cyclic $(r,\delta)$-LRCs with Length $q+1$ and its Factors}\label{q+1}

In this section, we give constructions of optimal cyclic $(r,\delta)$-LRCs with longer length $n\mid q+1$ based on Berlekamp-Justesen codes. For the ease of understanding, we firstly consider the case of $\delta=2$ or the optimal $q$-ary cyclic $r$-local LRCs in \ref{2016 ISIT}. Note that when $r\mid k$, the proposed constructions can be regarded as that given in \cite{Tamo cyclic} based on the $q$-ary subfield subcodes of Reed-Solomon codes with length $n\mid q^{2}-1$ over $\ff_{q^{2}}$.
However, the complete defining sets in our constructions have to be chosen specially and artfully, which are nontrivial and should be carefully discussed for even and odd $q$, respectively.
Then, the constructions are generalized to the cases of $(r, \delta)$-LRCs, and we obtain optimal $q$-ary cyclic $(r, \delta)$-LRCs ($\delta\ge 2$) in \ref{generalized q+1}, where more detailed discussion are needed for even and odd $\delta$, respectively. It is noticed that the optimal cyclic LRCs based on Berlekamp-Justesen codes could have longer length $q+1$ for a given alphabet with size $q$.

\subsection{Optimal Cyclic $r$-local LRCs with Length $q+1$ and its Factors}\label{2016 ISIT}

Just like the cyclic MDS codes with length $q+1$, this subsection has to be divided into two parts, even $q$ and odd $q$, because of the difference of structure of zeros.

\subsubsection{Cyclic optimal LRCs with even $q$}\label{q even}
In this subsection, assume that $n\mid q+1=2^{m}+1$, $r+1\mid n$,  and $r\mid k$. Let $ \nu=n/(r+1)$ and $\mu=k/r$. Therefore, $n$ is always an odd integer, while $r$  and $k$ are always even.

\bigskip
\textbf{Case 1:} $b$ is odd, e.g., $b=1$

\begin{theorem}\label{thm 4.1}
Let $\alpha \in \ff_{q^2}$ be a primitive $n$-th root of unity, where $n\mid q+1$. Let $b$ be a positive odd integer such that $(b, n) = 1$.
If $\mu=k/r$ is even, consider the following sets of elements of $\ff_{q^2}$:
\begin{eqnarray*}
D&=&\{\alpha^{\pm(\frac{n-b}{2})}, \alpha^{\pm(\frac{n-b}{2}-b)}, \alpha^{\pm(\frac{n-b}{2}-2b)},  \dots,  \alpha^{\pm(\frac{n-b}{2}-\frac{(n-k-\mu-1)}{2}b)} \},\\
L&=&\{\alpha^{i}\mid i \mod\, (r+1)=0\},
\end{eqnarray*}
then the cyclic code with the complete defining set of $L\cup D$ is an optimal $(n, k, r)$ $q$-ary cyclic $r$-local LRC.
\end{theorem}
\begin{IEEEproof}
For brevity, we only give proofs in the case of $b=1$, while the others are similar but need more detailed discussions.
Note that when $b=1$,
\begin{eqnarray*}
D=\{\alpha^{\pm(\frac{n-1}{2})}, \alpha^{\pm(\frac{n-3}{2})},\dots,  \alpha^{\pm\frac{1}{2}(k+\mu)} \}
\end{eqnarray*}
contains consecutive zeros since $\alpha^{-i}=\alpha^{n-i}$ and $k+\mu$ is even. Clearly, $|L|={n}/{(r+1)},\; |D|=n-k-\mu+1.$
So $\frac{1}{2}(k+\mu)=\frac{k}{2r}(r+1)$ is a multiple of $r+1$, which implies that
\begin{eqnarray*}
|L\cap D|
&=&2\times\left(\left\lfloor\frac{n-1}{2(r+1)}\right\rfloor-\frac{k}{2r}+1\right)\\
&=&\frac{n}{r+1}-\frac{k}{r}+1.
\end{eqnarray*}
Hence, $|L\cup D|=|L|+|D|-|L\cap D|=n-k$, which implies that the dimension of $\cC$ is $k$.
Locality follows by Lemma \ref{pro 2.3} for the set of zeros $L$, and the optimality follows by the BCH bound for the set of zeros $D$ and the  bound (\ref{singleton}). These complete the proof.
\end{IEEEproof}

{\em Remark 3:}
\begin{enumerate}
\item In Theorem \ref{thm 4.1}, the number of the consecutive zeros in $D$ is always even, so we choose paired zeros for $D$ (see Figure 3) based on the structure of zeros of $x^{n}-1$. Moreover, the remaining zeros of $g(x)$ are exactly in $L\backslash D$.
\item If $r\nmid k$, $k$ and $\lceil{{k}/{r}}\rceil$ are even, $\lceil{{k}/{r}}\rceil\ge4$, we can choose the zeros as Theorem \ref{thm 4.1} and obtain optimal $(n, k, r)$ LRCs. Because we also have $|D|=n-k-\lceil{{k}/{r}}\rceil+1$, $|L|=n/(r+1)$,  $|L\cap D|=\nu- \lceil{{k}/{r}}\rceil+1$, then $|L-L\cap D|=\lceil{{k}/{r}}\rceil-1>2$, which implies that there exist $r$ consecutive zeros with uniform distance $b$ in the complete defining set of $\cC^{\bot}$. Due to Proposition \ref{pro 2.3} and BCH bound, we obtain $d(\cC^{\bot})=r+1$, which means that the locality parameter is exactly $r$. The optimality also follows by the BCH bound for the set of zeros $D$ and the  bound (\ref{singleton}).
\end{enumerate}
\begin{center}
\includegraphics[width=2.4in]{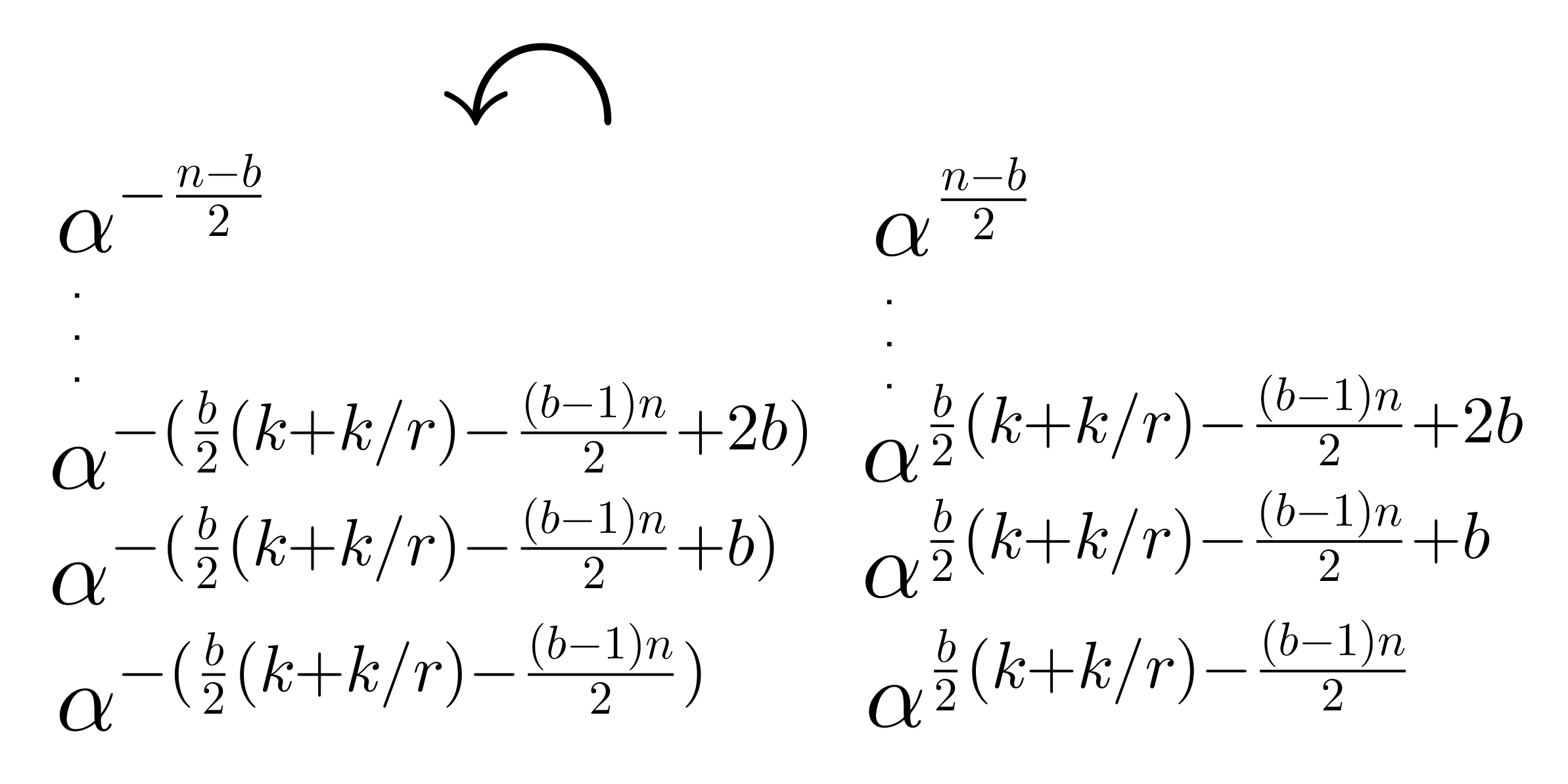}

{\small Fig.~3:  The consecutive zeros in $D$ for odd $b$ and even $\mu$.}
\end{center}






\begin{theorem}\label{thm 4.2}
Let $\alpha \in \ff_{q^2}$ be a primitive $n$-th root of unity, where $n\mid q+1$. Let $b$ be a positive odd integer such that $(b, n) = 1$.
If $\mu=k/r$ is odd, consider the following sets of elements of $\ff_{q^2}$:
 \begin{eqnarray*}
D&=&\{\alpha^{0},  \alpha^{\pm b}, \alpha^{\pm2b},\dots, \alpha^{\pm(\frac{n-k-k/r}{2})b}\},\\
L&=&\{\alpha^{i}\mid i \mod\, (r+1)=0\},
\end{eqnarray*}
then the cyclic code with the complete defining set of $L\cup D$ is an optimal $(n, k, r)$ q-ary cyclic $r$-local LRC.
\end{theorem}
\begin{IEEEproof}
For brevity, we only give proofs in the case of $b=1$, while the others are similar but need more detailed discussions.
Note that when $b=1$,
\begin{eqnarray*}
D=\{\alpha^{0}, \alpha^{\pm1}, \alpha^{\pm2},  \dots,   \alpha^{\pm\frac{n-k-k/r}{2}} \}
\end{eqnarray*}
contains consecutive zeros. Clearly, $|L|={n}/{(r+1)},\; |D|=n-k-k/r+1.$
Moreover,
\begin{eqnarray*}
|L\cap D|
&=&2\times\left\lfloor\frac{n-k-k/r}{2(r+1)}\right\rfloor+1\\
&=&\frac{n}{r+1}-k/r+1.
\end{eqnarray*}
Hence, $|L\cup D|=|L|+|D|-|L\cap D|=n-k$, which implies that the dimension of $\cC$ is $k$.
Locality and optimality follow by Lemma \ref{pro 2.3} and the BCH bound for the sets $L\cup D$.
\end{IEEEproof}
{\em Remark 4:}
\begin{enumerate}
\item In Theorem \ref{thm 4.2}, the number of the consecutive zeros in $D$ is always odd, so we choose paired zeros plus $\alpha^0$ for $D$ (see Figure 4) based on the structure of zeros of $x^{n}-1$. Moreover, the remaining zeros of $g(x)$ are exactly in $L\backslash D$ as well.
\item If $r\nmid k$, $k$ is even and  $\lceil{{k}/{r}}\rceil$ is odd. We can choose the zeros as Theorem \ref{thm 4.2} and obtain optimal $(n, k, r)$ LRCs. Because we also have  $|L-L\cap D|=\lceil{{k}/{r}}\rceil-1\ge2$, which implies that there exist $r$ consecutive zeros with uniform distance $b$ in the complete defining set of $\cC^{\bot}$, so $d(\cC^{\bot})=r+1$, which means that the locality parameter is exactly $r$. The optimality also follows by the BCH bound for the set of zeros $D$ and the  bound (\ref{singleton}).
\end{enumerate}

\begin{center}
\includegraphics[width=1.4in]{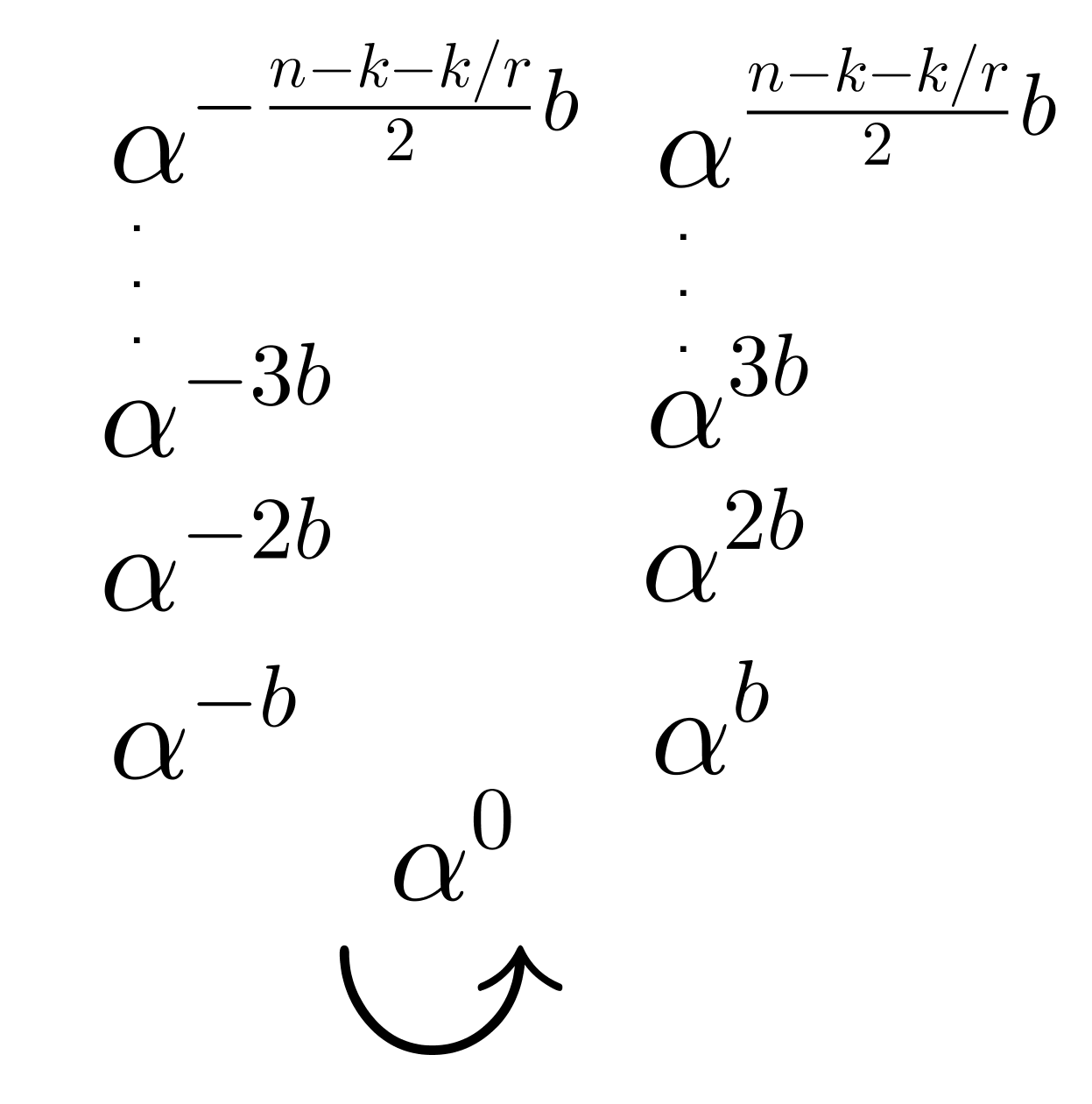}

{\small Fig.~4:  The consecutive zeros in $D$ for odd $b$ and odd $\mu$.}
\end{center}







\bigskip
\textbf{Case 2:} $b$ is even, e.g., $b=2$
\begin{theorem}\label{thm 4.3}
Let $\alpha \in \ff_{q^2}$ be a primitive $n$-th root of unity, where $n\mid q+1$. Let $b$ be a positive even integer such that $(b, n) = 1.$
If $\mu=k/r$ is even, consider the following sets of elements of $\ff_{q^2}$:
 \begin{eqnarray*}
D&=&\{\alpha^{\pm \frac{b}{2}},  \alpha^{\pm(\frac{b}{2}+b)}, \alpha^{\pm(\frac{b}{2}+2b)}, \dots, \alpha^{\pm(\frac{b}{2}+(\frac{n-k-k/r-1}{2})b)}\},\\
L&=&\{\alpha^{i}\mid i \mod\, (r+1)=0\},
\end{eqnarray*}
then the  cyclic code with the complete defining set of $L\cup D$ is an optimal $(n, k, r)$ q-ary cyclic $r$-local LRC.
\end{theorem}
The proof is similar to Theorem \ref{thm 4.2} and note that the consecutive zeros in $D$ is constructed as Figure 5.

\begin{center}
\includegraphics[width=1.8in]{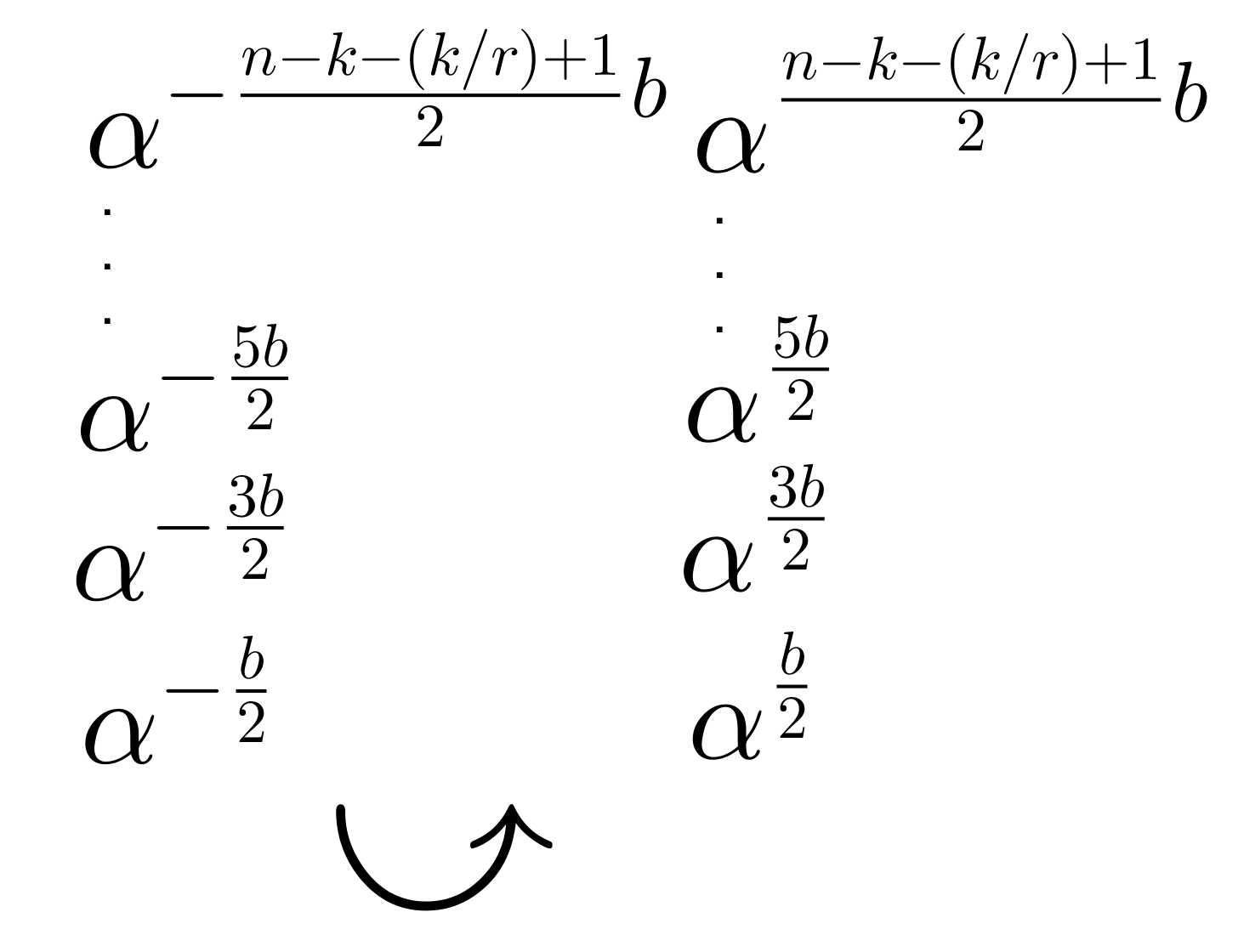}

{\small Fig.~5:  The consecutive zeros in $D$ for even $b$ and even $\mu$.}
\end{center}
{\em Remark 5:}
If $r\nmid k$, $k$ and $\lceil{{k}/{r}}\rceil$ are even, $\lceil{{k}/{r}}\rceil\ge4$. We can choose the zeros as Theorem \ref{thm 4.3} and obtain optimal $(n, k, r)$ LRCs. Because we also have  $|L-L\cap D|=\lceil{{k}/{r}}\rceil-1>2$, which implies that there exist $r$  consecutive zeros with uniform distance $b$ in the complete defining set of $\cC^{\bot}$, so $d(\cC^{\bot})=r+1$, which means that the locality parameter is exactly $r$. The optimality also follows by the BCH bound for the set of zeros $D$ and the  bound (\ref{singleton}).






\begin{theorem}\label{thm 4.4}
Let $\alpha \in \ff_{q^2}$ be a primitive $n$-th root of unity, where $n\mid q+1$. Let $b$ be a positive even integer such that $(b, n) = 1$.
If $\mu=k/r$ is odd, consider the following sets of elements of $\ff_{q^2}$:
 \begin{eqnarray*}
D&=&\{\alpha^{0},  \alpha^{\pm b}, \alpha^{\pm2b},\dots,\alpha^{\pm(\frac{n-k-k/r}{2})b} \},\\
L&=&\{\alpha^{i}\mid i \mod\, (r+1)=0\},
\end{eqnarray*}
then  the  cyclic code with the complete defining set of $L\cup D$ is an optimal $(n, k, r)$ q-ary cyclic $r$-local LRC.
\end{theorem}

The proof is similar to Theorem \ref{thm 4.2} and note that the consecutive zeros in $D$ is constructed as Figure 6.

\begin{center}
\includegraphics[width=1.4in]{figure2.jpg}

{\small Fig.~6: The consecutive zeros in $D$ for even $b$ and odd $\mu$.}
\end{center}

{\em Remark 6:}
If $r\nmid k$, $k$ is even and  $\lceil{{k}/{r}}\rceil$ is odd. We can choose the zeros as Theorem \ref{thm 4.4} and obtain optimal $(n, k, r)$ LRCs. Because we also have  $|L-L\cap D|=\lceil{{k}/{r}}\rceil-1\ge2$, which implies that there exist $r$  consecutive zeros with uniform distance $b$ in the complete defining set of $\cC^{\bot}$, so $d(\cC^{\bot})=r+1$, which means that the locality parameter is exactly $r$. The optimality also follows by the BCH bound for the set of zeros $D$ and the  bound (\ref{singleton}).




\bigskip
{\em Remark 7:}  In the above constructions, if we take $L=\{\alpha^{i}\mid i \mod (r+1)=l, \;0\le i\le n \}$ for some $0<l\le r$, the $L\cup D$ construction could not give any optimal cyclic LRCs. The reason is given as follows:

Firstly, $\alpha^{i}\in L$ implies $\alpha^{-i}\not\in L$. Assume the contrary that $\alpha^{-i}\in L$,
then $r+1 \mid i-l$ and $r+1 \mid -i-l$, which implies $r+1\mid 2l$. Since $r+1$ is odd,
$r+1\mid l$, which leads to a contradiction;

Hence, the structure of zeros of $x^{n}-1$ implies that the set
\begin{eqnarray*}
L'&=&\{\alpha^{-i}\mid \alpha^{i}\in L\}\\
&=&\{\alpha^{i}\mid i \mod (r+1)=r+1-l, \;0\le i\le n \}
\end{eqnarray*}
is contained in the complete defining set $L\cup D$ of $\cC$. Since $r+1$ is odd, $L\cap L'=\emptyset$, or $L'\subseteq D$. The zeros in $D$ are consecutive with uniform difference $b$, while zeros in $L'$ are consecutive with uniform difference $r+1$, so $b\mid r+1$. Moreover, since the $L\cup D$ construction requires $(b, n)=1$, $b=1$. At this time,
\begin{eqnarray*}
|D|&\ge& (|L'|-1)(r+1)+1\\
&=&\left(\frac{n}{r+1}-1\right)(r+1)+1\\
&=&n-r\\
&>&n-k-\mu+1,
\end{eqnarray*}
thus the $L\cup D$ construction could not give any optimal cyclic LRCs by the BCH bound.


\bigskip
\subsubsection{Cyclic optimal LRCs with odd $q$}
In this subsection, assume that $n\mid q+1$, $q$ is odd. $r+1\mid n$,  and $r\mid k$. Also let $ \nu=n/(r+1)$ and $\mu=k/r$. If $n$ is  an odd integer,  we can obtain optimal cyclic $r$-local LRCs as Theorems \ref{thm 4.1}- \ref{thm 4.4}. Therefore, we firstly give the following Theorem \ref{thm 4.5}-Theorem \ref{thm 4.7} of optimal $r$-local LRCs when $n\mid q+1$, $n$ is even. At last, Theorem \ref{thm 4.8} concludes the results in all cases.

\begin{theorem}\label{thm 4.5}
Let $\alpha \in \ff_{q^2}$ be a primitive $n$-th root of unity, where $n\mid q+1$, $n$ is even. Let $b$ be a positive odd integer such that $(b, n) = 1$.
If $\mu=k/r$ and $\nu={n}/{(r+1)}$  are odd, consider the following sets of elements of $\ff_{q^2}$:
 \begin{eqnarray*}
D&=&\{\alpha^{0},  \alpha^{\pm b}, \alpha^{\pm2b},\dots,\alpha^{\pm(\frac{n-k-k/r}{2})b} \},\\
L&=&\{\alpha^{i}\mid i \mod\, (r+1)=0\},
\end{eqnarray*}
then the cyclic code with the complete defining set of $L\cup D$ is an optimal $(n, k, r)$ q-ary cyclic $r$-local LRC.
\end{theorem}

The proof is similar to Theorem \ref{thm 4.2} and note that the consecutive zeros in $D$ is constructed as Figure 7.

\begin{center}
\includegraphics[width=1.4in]{figure2.jpg}

{\small Fig.~7:  The consecutive zeros in $D$ for odd $\mu$ and odd $\nu$.}
\end{center}
{\em Remark 8:}
If $r\nmid k$, $k$,  $\lceil{{k}/{r}}\rceil$ and $\nu$ are odd. We can choose the zeros as Theorem \ref{thm 4.5} and obtain optimal $(n, k, r)$ LRCs. Because we also have  $|L-L\cap D|=\lceil{{k}/{r}}\rceil-1\ge2$, which implies that there exist $r$ consecutive zeros with uniform distance $b$ in the complete defining set of $\cC^{\bot}$, so $d(\cC^{\bot})=r+1$, which means that the locality parameter is exactly $r$. The optimality also follows by the BCH bound for the set of zeros $D$ and the  bound (\ref{singleton}).

\begin{theorem}\label{thm 4.6}
Let $\alpha \in \ff_{q^2}$ be a primitive $n$-th root of unity, where $n\mid q+1$, $n$ is even. Let $b$ be a positive odd integer such that $(b, n) = 1$. If  $\mu=k/r$ is even, $\nu={n}/{(r+1)}$ is odd, consider the following sets of elements of $\ff_{q^2}$:
\begin{eqnarray*}
D&=&\{\alpha^{\frac{n}{2}},  \alpha^{\pm(\frac{n}{2}-b)}, \alpha^{\pm(\frac{n}{2}-2b)}, \dots,\alpha^{\pm(\frac{n}{2}-(\frac{n-k-k/r}{2})b)} \},\\
L&=&\{\alpha^{i}\mid i \mod\, (r+1)=0\},
\end{eqnarray*}
then  the  cyclic code with the complete defining set of $L\cup D$ is an optimal $(n, k, r)$ q-ary cyclic $r$-local LRC.
\end{theorem}

Note that $\nu={n}/{(r+1)}$ is odd, so $r+1\nmid {n}/{2}$, which implies $\alpha^{{n}/{2}}\notin L$. The proof is similar to Theorem \ref{thm 4.1}, the consecutive zeros in $D$ is constructed as Figure 8.
\begin{center}
\includegraphics[width=1.6in]{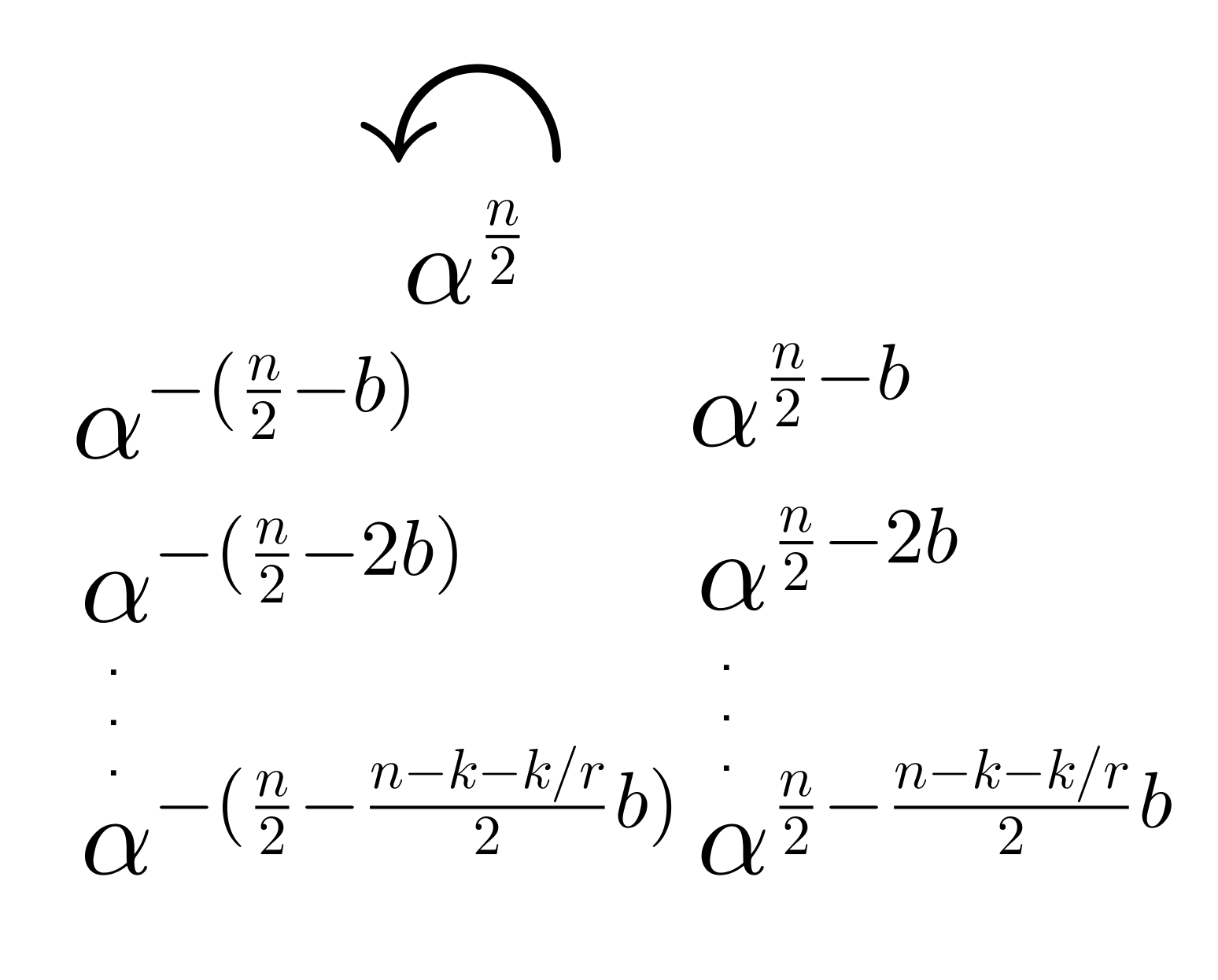}

{\small Fig.~8:  The consecutive zeros in $D$ for even $\mu$ and odd $\nu$.}
\end{center}
{\em Remark 9:}
If $r\nmid k$, $k$ and $\lceil{{k}/{r}}\rceil$ are even, $\nu$ is odd,  $\lceil{{k}/{r}}\rceil\ge4$. We can choose the zeros as Theorem \ref{thm 4.6} and obtain optimal $(n, k, r)$ LRCs. Because we also have  $|L-L\cap D|=\lceil{{k}/{r}}\rceil-1>2$, which implies that there exist $r$  consecutive zeros with uniform distance $b$ in the complete defining set of $\cC^{\bot}$, so $d(\cC^{\bot})=r+1$, which means that the locality parameter is exactly $r$. The optimality also follows by the BCH bound for the set of zeros $D$ and the  bound (\ref{singleton}).

\begin{theorem}\label{thm 4.7}
Let $\alpha \in \ff_{q^2}$ be a primitive $n$-th root of unity, where $n\mid q+1$, $n$ is even. Let $b$ be a positive odd integer such that $(b, n) = 1$. If  $\mu=k/r$ and $\nu={n}/{(r+1)}$ are even, consider the following sets of elements of $\ff_{q^2}$:
\begin{eqnarray*}
D&=&\{\alpha^{0},  \alpha^{\pm b}, \alpha^{\pm2b},\dots,\alpha^{\pm(\frac{n-k-k/r}{2})b} \},\\
L&=&\{\alpha^{i}\mid i \mod\, (r+1)=0\},
\end{eqnarray*}
\begin{eqnarray*}
\mbox{\rm or }\;\;D&=&\{\alpha^{\frac{n}{2}},  \alpha^{\pm(\frac{n}{2}-b)}, \alpha^{\pm(\frac{n}{2}-2b)}, \dots,\alpha^{\pm(\frac{n}{2}-(\frac{n-k-k/r}{2})b)} \},\\
L&=&\{\alpha^{i}\mid i \mod\, (r+1)=0\},
\end{eqnarray*}
then the cyclic codes with the complete defining sets of $L\cup D$ for both cases are optimal $(n, k, r)$ q-ary cyclic $r$-local LRCs.
\end{theorem}

Note that $\nu=n/(r+1)$ is even, so $r+1\mid n/2$, which implies $ \alpha^{{n}/{2}}\in L$. The proof is similar to Theorem  \ref{thm 4.1}$\thicksim$Theorem \ref{thm 4.2} and the consecutive zeros in $D$ for both cases are constructed as Figure 9.
\begin{center}
\includegraphics[width=2.8in]{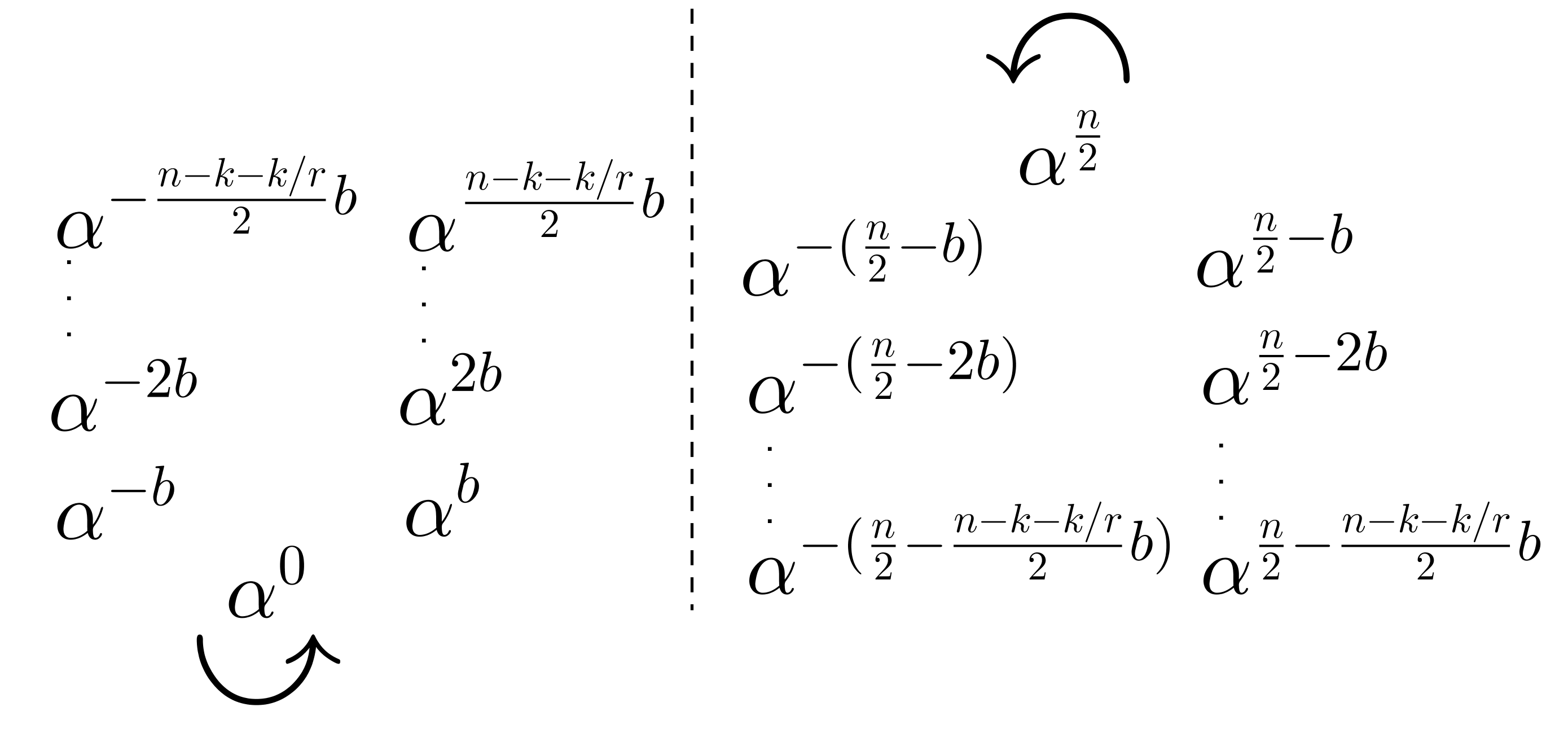}

{\small Fig.~9:  The consecutive zeros in $D$ for even $\mu$ and even $\nu$.}
\end{center}

{\em Remark 10:}
If $r\nmid k$, $k$, $\lceil{{k}/{r}}\rceil$ and  $\nu$ are even,  $\lceil{{k}/{r}}\rceil\ge4$. We can choose the zeros as Theorem \ref{thm 4.7} and obtain optimal $(n, k, r)$ LRCs. Because we also have  $|L-L\cap D|=\lceil{{k}/{r}}\rceil-1>2$, which implies that there exist $r$ consecutive zeros with uniform distance $b$ in the complete defining set of $\cC^{\bot}$, so $d(\cC^{\bot})=r+1$, which means that the locality parameter is exactly $r$. The optimality also follows by the BCH bound for the set of zeros $D$ and the  bound (\ref{singleton}).

Based on the above constructions, it is not difficult to obtain the following generalization.
\begin{theorem}\label{thm 4.8}
Let $\alpha \in \ff_{q^2}$ be a primitive $n$-th root of unity, where $n\mid q+1$ and $q$ is odd. Let $b$ be a positive integer such that  $(b, n) = 1$.
\begin{enumerate}
	\item If $n$ is odd, we can obtain optimal cyclic $r$-local LRCs as Theorems \ref{thm 4.1}- \ref{thm 4.4} ;
	\item If $n$ is even, we can  obtain optimal cyclic $r$-local LRCs as Theorems \ref{thm 4.5}- \ref{thm 4.7}.
\end{enumerate}
\end{theorem}
\begin{table}[!hbp]\label{table2}
\centering
\caption{\bf Optimal Cyclic $r$-local LRCs over $\ff_{q}$ with length $n\mid q+1$ }
\begin{tabular}{|c|c|}
\hline
 Conditions &Consecutive zeros set $D$  \\
\hline
$q$ even, $\mu$ even; &\multirow{3}{*}{$\lbrace \alpha^{i}| \frac{k+\mu}{2}\le i\le \frac{2n-k-\mu}{2}\rbrace$}\\
or $q$ odd,  $\mu$ even, &  \\
$n$ odd. ($b=1$)&\\
\hline
$q$ even, $\mu$ odd; &\multirow{3}{*}{$\lbrace \alpha^{i}| -\frac{n-k-\mu}{2}\le i\le\frac{n-k-\mu}{2}\rbrace$}\\\
 or $q$ odd, $\mu$ odd, &\\
 $n$ odd. ($b=1$)&\\
 \hline
$q$ even, $\mu$ even; &\multirow{3}{*}{$\lbrace \alpha^{2i+1}| -\frac{n-k-\mu+1}{2}\le i \le\frac{n-k-\mu-1}{2}\rbrace$} \\\
or $q$ odd, $\mu$ even,&\\
$n$ odd. ($b=2$) &\\
  \hline
$q$ even, $\mu$ odd; &\multirow{3}{*}{$\lbrace \alpha^{2i}| -\frac{n-k-\mu}{2}\le i\le\frac{n-k-\mu}{2}\rbrace$}\\\
or $q$ odd, $\mu$ odd,  &\\
$n$ odd. ($b=2$)&\\
   \hline
$q$ odd, $\mu$ and $\nu$ odd, &\multirow{2}{*}{$\lbrace \alpha^{i}| -\frac{n-k-\mu}{2}\le i\le\frac{n-k-\mu}{2}\rbrace$} \\\
$n$ even. ($b=1$)&\\
   \hline
$q$ odd, $\mu$ even, &\multirow{2}{*}{$\lbrace \alpha^{i}| \frac{k+\mu}{2}\le i\le \frac{2n-k-\mu}{2}\rbrace$}\\\
 $\nu$ odd, $n$ even. ($b=1$)&\\
 \hline
&\multirow{2}{*}{$\lbrace  \alpha^{i}| -\frac{n-k-\mu}{2}\le i\le\frac{n-k-\mu}{2}\rbrace$}\\\
$q$ odd, $\mu$ and $\nu$ even,  &\\
 \cline{2-2}
 &\multirow{2}{*}{$\lbrace \alpha^{i}| \frac{k+\mu}{2}\le i\le \frac{2n-k-\mu}{2}\rbrace$}\\\
$n$ even. ($b=1$)  &\\
 \hline
\end{tabular}
\end{table}
\bigskip



\bigskip
\subsection{Optimal cyclic  $(r, \delta)$-LRCs with length $q+1$ and its factors}\label{generalized q+1}
Along with the ideas of optimal $r$-local LRCs with length $n\mid q+1$ in Section \ref{2016 ISIT}, we can naturally obtain the corresponding generalized constructions of  optimal cyclic $q$-ary $(r, \delta)$-LRCs with length $n\mid q+1$  in this subsection.
We assume that $r+\delta-1\mid n$, $r\mid k$ and let $\nu^{\prime}=n/(r+\delta-1)$, $\mu=k/r$. As for $r\nmid k$, we can also obtain optimal cyclic $(r, \delta)$-LRCs as Subsection \ref{2016 ISIT}, which need more detailed discussions as Remark 3 $\thicksim$ Remark 10.
\subsubsection{Optimal cyclic $(r, \delta)$-LRCs with even $\delta$}
We firstly consider the constructions of  optimal cyclic  $(r, \delta)$-LRCs with even $\delta$, which are generalized constructions corresponding to those in Subsection  \ref{2016 ISIT}. Based on the  \emph{$(\cup_{m}L_{m})\cup D$ construction} and even $\delta$, we have to choose pairs of  $(r, \delta)$-locality sets $L_{\pm m}$ plus $L_0$.

Similar to  Lemma \ref{lemma 4.1},  if $\alpha \in \ff_{q^{2}}$ is a primitive $n$-th root of unity, we have the following lemma:
 \begin{lemma}\label{lemma:z3} Let $i_{0}=0<i_{1}<\dots <i_{(\delta-2)/2}\le r+\delta-2$
be an arithmetic progression with $\delta/2$ items and common difference $b$
such that $(b,n)=1.$ And consider a $(\delta-1)\nu^{\prime}\times n$ matrix $\cH$ with the rows
\begin{eqnarray*}
  h_{m}^{(j)}&=&(1, \alpha^{m(r+\delta-1)-i_{|j|}},\alpha^{2(m(r+\delta-1)-i_{|j|})}, \dots,\alpha^{(n-1)(m(r+\delta-1)-i_{|j|})}), \quad j<0,\\
  h_{m}^{(j)}&=&(1, \alpha^{m(r+\delta-1)+i_{j}},\alpha^{2(m(r+\delta-1)+i_{j})}, \dots,\alpha^{(n-1)(m(r+\delta-1)+i_{j})}), \quad j\ge0,
\end{eqnarray*}
 where $j=0, \pm1, \dots, \pm((\delta-2)/2-1), \pm (\delta-2)/2$, $m=0,1,\dots,\nu^{\prime}-1,$ and $\nu^{\prime}=n/(r+\delta-1)$.
Then  all the cyclic shifts of the row vectors $v_j$ of weight $r+\delta-1$ in the  $(\delta-1)\times n$-matrix $V=\big(v_j)_{(\delta-1)\times n}$
are contained in the row space of $\cH$ over  $\ff_{q^2}$,
where
   $$
  v_j=
 \begin{pmatrix}
1\underbrace{0 \ldots 0}_{\nu^{\prime}-1}&\alpha^{-i_{|j|}\nu^{\prime}}&\underbrace{0\ldots0}_{\nu^{\prime}-1}&\alpha^{-2i_{|j|}\nu^{\prime}}&\ldots
&\alpha^{-(r+\delta-2)i_{|j|}\nu^{\prime}}&\underbrace{0\ldots0}_{\nu^{\prime}-1}
\end{pmatrix}
 , j<0,
  $$

     $$
  v_j=
 \begin{pmatrix}
1\underbrace{0 \ldots 0}_{\nu^{\prime}-1}&\alpha^{i_{j}\nu^{\prime}}&\underbrace{0\ldots0}_{\nu^{\prime}-1}&\alpha^{2i_{j}\nu^{\prime}}&\ldots
&\alpha^{(r+\delta-2)i_{j}\nu^{\prime}}&\underbrace{0\ldots0}_{\nu^{\prime}-1}\\
\end{pmatrix}
, j\ge 0,
  $$
  $j=0, \pm1, \dots, \pm((\delta-2)/2-1), \pm (\delta-2)/2$, $m=0,1,\dots,\nu^{\prime}-1,$ and $\nu^{\prime}=n/(r+\delta-1)$.
\end{lemma}
\begin{IEEEproof}
The proof of this lemma is similar to Lemma \ref{lemma 4.1}, the only difference is that when $n\mid q+1$, the order of $q$ modulo $n$ is $2$, which implies that the splitting field of $x^{n} -1$ is $\ff_{q^{2}}$. Similar to Lemma \ref{lem 2.2}, we can obtain that for any fixed $j\in\{0, \pm1, \dots, \pm((\delta-2)/2-1), \pm (\delta-2)/2\}$,  all the cyclic shifts of the $n$-dimensional row vectors $v_{j}$ of weight $r+\delta-1$ are contained in the row space of $\cH$ over  $\ff_{q^2}$.
\end{IEEEproof}

We  denote $[\alpha^{i}]$ as the column vector in $\ff_{q}^{2}$ corresponding to the element $\alpha^{i}$ in $\ff_{q^2}$, then the matrix $V$ corresponds to an $2(\delta-1)\times n$ matrix over  $\ff_q$. The non-zero columns  correspond to an $2(\delta-1)\times (r+\delta-1)$ matrix $V^{\prime}$ over  $\ff_q$:
$$
 \begin{pmatrix}
[1]&[\alpha^{-i_{(\delta-2)/2}\nu^{\prime}}]&[(\alpha^{-i_{(\delta-2)/2}\nu^{\prime}})^{2}]&\ldots
&[(\alpha^{-i_{(\delta-2)/2}\nu^{\prime}})^{r+\delta-2}]\\
\vdots&\vdots&\vdots&\vdots&\vdots\\
[1]&[1]&[1]&\ldots
&[1]\\
\vdots&\vdots&\vdots&\vdots&\vdots\\
[1]&[\alpha^{i_{(\delta-2)/2}\nu^{\prime}}]&[(\alpha^{i_{(\delta-2)/2}\nu^{\prime}})^{2}]&\ldots
&[(\alpha^{i_{(\delta-2)/2}\nu^{\prime}})^{r+\delta-2}]\\
\end{pmatrix}
  $$
Since any $\delta-1$ columns of $V^{\prime}$ are linearly independent over $\ff_{q^2}$, any $\delta-1$ columns of $V^{\prime}$ are linearly independent over $\ff_{q}$.  And all the cyclic shifts of the row vectors of weight $r+\delta-1$ in $V$  are contained in the row space of $\cH$ over  $\ff_{q^2}$, then  the cyclic shifts of the row vectors in $V$ partition the support of the code into disjoint subsets of size $r+\delta-1$ which define the $(r, \delta)$-local recovering sets of the symbols. Therefore we obtain the following proposition.
\begin{proposition} \label{prop: 4.5}
Let $\ff_{q^s}$ be the splitting field of $x^n-1$, where $n\mid q+1$, let $\cC$ be a cyclic code of length $n$ over $\ff_q$ with the complete defining set $Z$, and
let $r$, $\delta$ be positive integers such that $(r+\delta-1)\mid n.$
Let $i_{0}=0<i_{1}<\dots <i_{(\delta-2)/2}\le r+\delta-2$
be an arithmetic progression with $\delta/2$ items and common difference $b$
such that $(b,n)=1.$
If $Z$ contains some cosets of the group of $\nu^{\prime}$-th roots of unity $\cup_{m}L_{m}$,
where
$$L_{m}=\{\alpha^i\mid i \mod (r+\delta-1)=m\}, \, m=i_{0}, \pm i_{1}, \pm i_{2}, \dots, \pm i_{(\delta-2)/2},$$
then $\cC$ has $(r, \delta)$-locality.
\end{proposition}

In order to simplify the discussion, we choose $b=1$ if $b$ is odd such that $(b, n)=1$, and $b=2$ if $b$ is even such that $(b, n)=1$ in the following constructions, respectively.
\begin{cnstr}\label{construction 2}
Let $\alpha \in \ff_{q^2}$ be a primitive $n$-th root of unity, where $n\mid q+1$ and $q$ is even. If $\mu=k/r$ is even, consider the following sets of elements of $\ff_{q^2}$:
\begin{eqnarray*}
D&=&\{\alpha^{\pm(\frac{n-1}{2})}, \alpha^{\pm(\frac{n-1}{2}-1)}, \alpha^{\pm(\frac{n-1}{2}-2)}, \dots,  \alpha^{\pm(\frac{n-1}{2}-\frac{n-k-(\mu-1)(\delta-1)-2}{2})} \},\\
L_{m}&=&\{\alpha^{i}\mid i \mod\, (r+\delta-1)=m\}, m=0, \pm1, \pm2, \dots, \pm(\delta-2)/2.
\end{eqnarray*}
Then the cyclic code $\cC$ with the complete defining set of zeros $(\cup_{m}L_{m})\cup D$ is an optimal $q$-ary cyclic $(r, \delta)$-LRC.
\end{cnstr}
\begin{IEEEproof}
From the construction above, we have:
$|D|=n-k-(\mu-1)(\delta-1)$,
$|L_{m}|=n/(r+\delta-1)$, $m=0, \pm1, \pm2, \dots, \pm(\delta-2)/2$. Since $k/r$ is even,
$\frac{n-1}{2}-\frac{n-k-(\mu-1)(\delta-1)-2}{2}=\frac{k+(\mu-1)(\delta-1)+1}{2}=\frac{k}{2r}(r+\delta-1)-\frac{\delta-2}{2}$, which implies that
\begin{eqnarray*}
\frac{n-1}{2}-\frac{n-k-(\mu-1)(\delta-1)-2}{2} \mod\, (r+\delta-1)=-(\delta-2)/2,
\end{eqnarray*}
\begin{eqnarray*}
-\left(\frac{n-1}{2}-\frac{n-k-(\mu-1)(\delta-1)-2}{2}\right) \mod\, (r+\delta-1)=(\delta-2)/2.
\end{eqnarray*}
Then for any $L_{m}$, $m=0, \pm1, \pm2, \dots, \pm(\delta-2)/2$,
\begin{eqnarray*}
|L_{m}\cap D|
&=&\left(\left\lfloor\frac{n-k-(\mu-1)(\delta-1)}{r+\delta-1}\right\rfloor+1\right)\\
&=&\left(\left\lfloor\frac{n-\mu(r+\delta-1)+\delta-2}{r+\delta-1}\right\rfloor+1\right)\\
&=& \nu^{\prime}-\mu+1.
\end{eqnarray*}
Hence, $|(\cup_{m}L)\cup D|=(\delta-1)|L_{m}|+|D|-(\delta-1)|L_{m}\cap D|=n-k$, which implies that the dimension of $\cC$ is $k$.
$(r, \delta)$-Locality follows by Proposition \ref{prop: 4.5}  for the sets of zeros $\cup L_{m}$, and the optimality follows by the BCH bound for the set of zeros $D$ and the generalized Singleton-like bound (\ref{eq_GeneralizedSingleton}). These complete the proof.
\end{IEEEproof}

Example \ref{ex 3} below can help understand the construction above.  In the rest of paper, the bold numbers in $L_{m}$ means that they lies in $D$.
\begin{xmpl}\label{ex 3}
Let $n=q+1=2^{6}+1=65=5\times 13$, $r=2$, $\delta=4$, $k=12$, $b=1$. We have
\begin{eqnarray*}
D&=&\{ \alpha^{i}\mid i=\pm14, \pm15, \pm16, \dots, \pm32\}, \\
L_{0}&=&\{\alpha^{i}\mid i \mod\, 5=0\}\\
&=&\{\alpha^{i}\mid i=0, \pm5, \pm10, \bm{\pm15, \pm 20, \pm25, \pm30} \},\\
L_{1}&=&\{\alpha^{i}\mid i \mod\, 5=1\}\\
&=&\{\alpha^{i}\mid i=1, 6, 11, \bm{16, 21, 26, 31, -29, -24,} \bm{ -19, -14,}-9, -4 \},\\
L_{-1}&=&\{\alpha^{i}\mid i \mod\, 5=-1\}\\
&=&\{\alpha^{i}\mid i=4, 9, \bm{14, 19, 24, 29, -31, -26,} \bm{ -21, -16,} -11, -6, -1 \}.
\end{eqnarray*}
Then the cyclic code $\cC$ with the complete defining set of zeros $(L_{-1}\cup L_{0}\cup L_{1})\cup D$ is an optimal $64$-ary cyclic $(2, 4)$-LRC.
\end{xmpl}

\begin{cnstr}\label{construction 3}
Let $\alpha \in \ff_{q^2}$ be a primitive $n$-th root of unity, where $n\mid q+1$ and $q$ is even.
If $\mu=k/r$ is odd, consider the following sets of elements of $\ff_{q^2}$:
\begin{eqnarray*}
D&=&\{\alpha^{0},  \alpha^{\pm 1}, \alpha^{\pm2},\dots, \alpha^{\pm(\frac{n-k-(\mu-1)(\delta-1)-1}{2})}\},\\
L_{m}&=&\{\alpha^{i}\mid i \mod\, (r+\delta-1)=m\}, m=0, \pm1, \pm2, \dots, \pm(\delta-2)/2.
\end{eqnarray*}
Then the cyclic code $\cC$ with the complete defining set of zeros $(\cup_{m}L_{m})\cup D$ is an optimal $q$-ary cyclic $(r, \delta)$-LRC.
\end{cnstr}
\begin{IEEEproof}
Clearly, we have $|D|=n-k-(\mu-1)(\delta-1)$, $|L_{m}|=n/(r+\delta-1), \, m=0, \pm1, \pm2, \dots, \pm(\delta-2)/2$.
Since $\mu=k/r$ is odd, $\frac{n-k-(\mu-1)(\delta-1)-1}{2}=\frac{\nu^{\prime}-\mu}{2}(r+\delta-1)+\frac{\delta-2}{2}$, which implies:
\begin{eqnarray*}
&\quad&\frac{n-k-(\mu-1)(\delta-1)-1}{2}\mod\, (r+\delta-1)=(\delta-2)/2;
\end{eqnarray*}
\begin{eqnarray*}
&\quad&-\frac{n-k-(\mu-1)(\delta-1)-1}{2}\mod\, (r+\delta-1)=-(\delta-2)/2.
\end{eqnarray*}
Then for any $L_{m}$, $m=0, \pm1, \pm2, \dots, \pm(\delta-2)/2$,
\begin{eqnarray*}
|L_{m}\cap D|
&=&\left(\left\lfloor\frac{n-k-(\mu-1)(\delta-1)}{r+\delta-1}\right\rfloor+1\right)\\
&=& \nu^{\prime}-\mu+1.
\end{eqnarray*}
Hence, $|(\cup_{m}L)\cup D|=n-k$, which implies that the dimension of $\cC$ is $k$.
$(r, \delta)$-Locality follows by Proposition \ref{prop: 4.5}  for the sets of zeros $\cup L_{m}$, and the optimality follows by the BCH bound for the set of zeros $D$ and the generalized Singleton-like bound (\ref{eq_GeneralizedSingleton}). These complete the proof.
\end{IEEEproof}
\begin{xmpl}\label{ex 4}
Let $n=q+1=2^{6}+1=65=5\times 13$, $r=2$, $\delta=4$, $k=14$, $b=1$. We have
\begin{eqnarray*}
D&=&\{ \alpha^{i}\mid i=0, \pm1, \pm2, \pm3 \dots, \pm16\}, \\
L_{0}&=&\{\alpha^{i}\mid i \mod\, 5=0\}\\
&=&\{\alpha^{i}\mid i=\bm{0, \pm5, \pm10, \pm15,} \pm 20, \pm25, \pm30 \},\\
L_{1}&=&\{\alpha^{i}\mid i \mod\, 5=1\}\\
&=&\{\alpha^{i}\mid i=\bm{1, 6, 11, 16,} 21, 26, 31, -29, -24, -19, \bm{ -14, -9, -4} \},\\
L_{-1}&=&\{\alpha^{i}\mid i \mod\, 5=-1\}\\
&=&\{\alpha^{i}\mid i=\bm{4, 9, 14,} 19, 24, 29, -31, -26,  -21, \bm{-16, -11, -6, -1} \}.
\end{eqnarray*}
Then the cyclic code $\cC$ with the complete defining set of zeros $(L_{-1}\cup L_{0}\cup L_{1})\cup D$ is an optimal $64$-ary cyclic $(2, 4)$-LRC.
\end{xmpl}
{\em Remark 11:}
\begin{enumerate}
\item  As mentioned previously, in Construction \ref{construction 2} and Construction \ref{construction 3}, we can choose the $(r, \delta)$-locality sets as some cosets of the group of $\nu^{\prime}$-th roots of unity, where  $m=i_{0}, \pm i_{1}, \pm i_{2}, \dots, \pm i_{(\delta-2)/2}$ and  $i_{0}=0<i_{1}<\dots <i_{(\delta-2)/2}\le r+\delta-2$ with uniform odd difference $b$ such that $(b,n)=1$, choose  the  consecutive zeros for $D$ with the same uniform difference $b$ in the similar way as Theorem \ref{thm 4.1} $\thicksim $\ref{thm 4.2} in Section \ref{2016 ISIT},, then we can obtain the more general constructions.
\item When $\delta=2$, the general construction stated above reduces to Theorem \ref{thm 4.1} and Theorem \ref{thm 4.2} in Section \ref{2016 ISIT}.
\end{enumerate}

\begin{cnstr}\label{construction 4}
Let $\alpha \in \ff_{q^2}$ be a primitive $n$-th root of unity, where $n\mid q+1$ and $q$ is even.
If $\mu=k/r$ is even, consider the following sets of elements of $\ff_{q^2}$:
\begin{eqnarray*}
D&=&\{\alpha^{\pm 1},  \alpha^{\pm3}, \alpha^{\pm5}, \dots, \alpha^{\pm(1+\frac{n-k-(\mu-1)(\delta-1)-2}{2}\times2)}\},\\
L_{m}&=&\{\alpha^{i}\mid i \mod\, (r+\delta-1)=m\}, m=0, \pm2, \pm4, \dots, \pm(\delta-2).
\end{eqnarray*}
Then the cyclic code $\cC$ with the defining set of zeros $(\cup_{m}L_{m})\cup D$ is an optimal $q$-ary cyclic $(r, \delta)$-LRC.
\end{cnstr}

Note that $1+\frac{n-k-(\mu-1)(\delta-1)-2}{2}\times2=n-k-(\mu-1)(\delta-1)-1=n-\mu(r+\delta-1)+\delta-2$, which implies
\begin{eqnarray*}
&\quad&1+\frac{n-k-(\mu-1)(\delta-1)-2}{2}\times2\mod\, (r+\delta-1)=\delta-2;
\end{eqnarray*}
\begin{eqnarray*}
&\quad&-\left(1+\frac{n-k-(\mu-1)(\delta-1)-2}{2}\times2\right)\mod\, (r+\delta-1)=-(\delta-2).
\end{eqnarray*}
The rest of the proof  is similar to Construction \ref{construction 2}.
\begin{xmpl}\label{ex 5}
Let $n=q+1=2^{6}+1=65=5\times 13$, $r=2$, $\delta=4$, $k=16$, $b=2$. We have
\begin{eqnarray*}
D&=&\{ \alpha^{i}\mid i= \pm1, \pm3, \pm5 \dots, \pm27\}, \\
L_{0}&=&\{\alpha^{i}\mid i \mod\, 5=0\}\\
&=&\{\alpha^{i}\mid i=0, \bm{\pm5,} \pm10, \bm{\pm15,} \pm 20, \bm{\pm25,} \pm30 \},\\
L_{2}&=&\{\alpha^{i}\mid i \mod\, 5=2\}\\
&=&\{\alpha^{i}\mid i=2,  \bm{7}, 12, \bm{17,} 22, \bm{27,} 32, -28, \bm{-23,} -18, \bm{ -13,} -8, \bm{-3} \},\\
L_{-2}&=&\{\alpha^{i}\mid i \mod\, 5=-2\}\\
&=&\{\alpha^{i}\mid i=\bm{3,} 8, \bm{13,} 18, \bm{23,} 28, -32, \bm{-27,} -22, \bm{-17,} -12, \bm{-7,} -2 \}.
\end{eqnarray*}
Then the cyclic code $\cC$ with the defining set of zeros $(L_{2}\cup L_{0}\cup L_{-2})\cup D$ is an optimal $64$-ary cyclic $(2, 4)$-LRC.
\end{xmpl}

\begin{cnstr}\label{construction 5}
Let $\alpha \in \ff_{q^2}$ be a primitive $n$-th root of unity,
where $n\mid q+1$ and $q$ is even.
If $\mu=k/r$ is odd, consider the following sets of elements of $\ff_{q^2}$:
\begin{eqnarray*}
D&=&\{\alpha^{0},  \alpha^{\pm 2}, \alpha^{\pm4},\dots, \alpha^{\pm(\frac{n-k-(\mu-1)(\delta-1)-1}{2})\times 2}\},\\
L_{m}&=&\{\alpha^{i}\mid i \mod\, (r+\delta-1)=m\}, m=0, \pm2, \pm4, \dots, \pm(\delta-2).
\end{eqnarray*}
Then the cyclic code $\cC$ with the defining set of zeros $(\cup_{m}L_{m})\cup D$ is an optimal $q$-ary cyclic $(r, \delta)$-LRC.
\end{cnstr}

The proof of this construction is similar to Construction \ref{construction 4}.
\begin{xmpl}\label{ex 6}
Let $n=q+1=2^{6}+1=65=5\times 13$, $r=2$, $\delta=4$, $k=18$, $b=2$. We have
\begin{eqnarray*}
D&=&\{ \alpha^{i}\mid i=0, \pm2, \pm4, \pm6 \dots, \pm22\}, \\
L_{0}&=&\{\alpha^{i}\mid i \mod\, 5=0\}\\
&=&\{\alpha^{i}\mid i=\bm{0,} \bm{\pm5,} \pm10, \bm{\pm15,} \pm 20, \pm25, \pm30 \},\\
L_{2}&=&\{\alpha^{i}\mid i \mod\, 5=2\}\\
&=&\{\alpha^{i}\mid i=\bm{2,}  7, \bm{12,}  17, \bm{22,} 27, 32, -28, -23, \bm{-18,} -13, \bm{-8,} -3 \},\\
L_{-2}&=&\{\alpha^{i}\mid i \mod\, 5=-2\}\\
&=&\{\alpha^{i}\mid i=3, \bm{8,} 13, \bm{18,} 23, 28, -32, -27,  \bm{-22,} -17, \bm{-12,} -7, \bm{-2} \}.
\end{eqnarray*}
Then the cyclic code $\cC$ with the defining set of zeros $(L_{2}\cup L_{0}\cup L_{-2})\cup D$ is an optimal $64$-ary cyclic $(2, 4)$-LRC.
\end{xmpl}
{\em Remark 12:}
\begin{enumerate}
\item  As mentioned previously, in Construction \ref{construction 4} and Construction \ref{construction 5}, we can choose the $(r, \delta)$-locality sets as some cosets of the group of $\nu^{\prime}$-th roots of unity, where  $m=i_{0}, \pm i_{1}, \pm i_{2}, \dots, \pm i_{(\delta-2)/2}$ and  $i_{0}=0<i_{1}<\dots <i_{(\delta-2)/2}\le r+\delta-2$ with uniform even difference $b$ such that $(b,n)=1$, choose  the  consecutive zeros for $D$ with the same uniform difference $b$ in the similar way as Theorem \ref{thm 4.3} $\thicksim $\ref{thm 4.4} in Section \ref{2016 ISIT}, then we can obtain the more general constructions.
\item When $\delta=2$, the general constructions stated above reduce to Theorem \ref{thm 4.3}  and Theorem \ref{thm 4.4} in Section \ref{2016 ISIT}.
\end{enumerate}
\begin{cnstr}\label{construction 6}
Let $\alpha \in \ff_{q^2}$ be a primitive $n$-th root of unity, where $n\mid q+1$, $q$ is odd and $n$ is even.
If $\mu=k/r$ and $\nu^{\prime}$  are odd, consider the following sets of elements of $\ff_{q^2}$:
\begin{eqnarray*}
D&=&\{\alpha^{0},  \alpha^{\pm 1}, \alpha^{\pm2},\dots, \alpha^{\pm(\frac{n-k-(\mu-1)(\delta-1)-1}{2})}\},\\
L_{m}&=&\{\alpha^{i}\mid i \mod\, (r+\delta-1)=m\}, m=0, \pm1, \pm2, \dots, \pm(\delta-2)/2.
\end{eqnarray*}
Then the cyclic code $\cC$ with the defining set of zeros $(\cup_{m}L_{m})\cup D$ is an optimal $q$-ary cyclic $(r, \delta)$-LRC.
\end{cnstr}

The proof of this construction is similar to Construction \ref{construction 3}.
\begin{xmpl}\label{ex 7}
Let $n=q+1=7^{2}+1=50=10\times 5,$ $r=5,$ $\delta=6,$ $k=15,$ $b=1$. We have
\begin{eqnarray*}
D&=&\{ \alpha^{i}\mid i=0, \pm1, \pm2, \pm3 \dots, \pm12\}, \\
L_{0}&=&\{\alpha^{i}\mid i \mod\, 10=0\}\\
&=&\{\alpha^{i}\mid i=\bm{0, \pm10,} \pm20 \},\\
L_{1}&=&\{\alpha^{i}\mid i \mod\, 10=1\}\\
&=&\{\alpha^{i}\mid i=\bm{1, 11,} 21, -19 \bm{-9} \},\\
L_{-1}&=&\{\alpha^{i}\mid i \mod\, 10=-1\}\\
&=&\{\alpha^{i}\mid i=\bm{9,} 19, -21, \bm{-11, -1} \},\\
L_{2}&=&\{\alpha^{i}\mid i \mod\, 10=2\}\\
&=&\{\alpha^{i}\mid i=\bm{2, 12,} 22, -18 \bm{-8} \},\\
L_{-2}&=&\{\alpha^{i}\mid i \mod\, 10=-2\}\\
&=&\{\alpha^{i}\mid i=\bm{8,} 18, -22, \bm{-12, -2} \}.
\end{eqnarray*}
Then the cyclic code $\cC$ with the defining set of zeros $( L_{-2}\cup L_{-1}\cup L_{0}\cup L_{1}\cup L_{2})\cup D$ is an optimal $49$-ary cyclic $(5, 6)$-LRC.
\end{xmpl}

\begin{cnstr}\label{construction 7}
Let $\alpha \in \ff_{q^2}$ be a primitive $n$-th root of unity, where $n\mid q+1$, $q$ is odd and $n$ is even.
If $\mu=k/r$ is even, $\nu^{\prime}$  is odd, consider the following sets of elements of $\ff_{q^2}$:
\begin{eqnarray*}
D&=&\{\alpha^{\frac{n}{2}},  \alpha^{\pm(\frac{n}{2}-1)}, \alpha^{\pm(\frac{n}{2}-2)}, \dots,\alpha^{\pm(\frac{n}{2}-(\frac{n-k-(\mu-1)(\delta-1)-1}{2}))} \},\\
L_{m}&=&\{\alpha^{i}\mid i \mod\, (r+\delta-1)=m\}, m=0, \pm1, \pm2, \dots, \pm(\delta-2)/2.
\end{eqnarray*}
Then the cyclic code $\cC$ with the defining set of zeros $(\cup_{m}L_{m})\cup D$ is an optimal $q$-ary cyclic $(r, \delta)$-LRC.
\end{cnstr}

The proof of this construction is similar to Construction \ref{construction 2}.
\begin{xmpl}\label{ex 8}
Let $n=q+1=7^{2}+1=50=10\times 5,$ $r=7,$ $\delta=4,$ $k=28,$ $b=1$. We have
\begin{eqnarray*}
D&=&\{ \alpha^{i}\mid i=\pm19, \pm20, \pm21 \dots, \pm24, 25\}, \\
L_{0}&=&\{\alpha^{i}\mid i \mod\, 10=0\}\\
&=&\{\alpha^{i}\mid i=0, \pm10, \bm{ \pm20} \},\\
L_{1}&=&\{\alpha^{i}\mid i \mod\, 10=1\}\\
&=&\{\alpha^{i}\mid i=1, 11,\bm{ 21, -19,} -9 \},\\
L_{-1}&=&\{\alpha^{i}\mid i \mod\, 10=-1\}\\
&=&\{\alpha^{i}\mid i=9, \bm{19, -21,} -11, -1 \}.
\end{eqnarray*}
Then the cyclic code $\cC$ with the defining set of zeros $(L_{1}\cup L_{0}\cup L_{-1})\cup D$ is an optimal $49$-ary cyclic $(7, 4)$-LRC.
\end{xmpl}

\begin{cnstr}\label{construction 8}
Let $\alpha \in \ff_{q^2}$ be a primitive $n$-th root of unity,
where $n\mid q+1$ , $q$ is odd and $n$ is even. If $k$, $\mu=k/r$ and $\nu^{\prime}$  are even, consider the following sets of elements of $\ff_{q^2}$:
\begin{eqnarray*}
D&=&\{\alpha^{0},  \alpha^{\pm 1}, \alpha^{\pm2},\dots, \alpha^{\pm(\frac{n-k-(\mu-1)(\delta-1)-1}{2})}\},\\
L_{m}&=&\{\alpha^{i}\mid i \mod\, (r+\delta-1)=m\}, m=0, \pm1, \pm2, \dots, \pm(\delta-2)/2.
\end{eqnarray*}
\begin{eqnarray*}
\mbox{\rm or }\;\;D&=&\{\alpha^{\frac{n}{2}},  \alpha^{\pm(\frac{n}{2}-1)}, \alpha^{\pm(\frac{n}{2}-2)}, \dots,\alpha^{\pm(\frac{n}{2}-(\frac{n-k-(\mu-1)(\delta-1)-1}{2}))} \},\\
L_{m}&=&\{\alpha^{i}\mid i \mod\, (r+\delta-1)=m\}, m=0, \pm1, \pm2, \dots, \pm(\delta-2)/2.
\end{eqnarray*}
Then the cyclic code $\cC$ with the defining set of zeros $(\cup_{m}L_{m})\cup D$ is an optimal $q$-ary cyclic $(r, \delta)$-LRC.
\end{cnstr}

The proof is similar to Construction \ref{construction 2} and Construction \ref{construction 3}.
\begin{xmpl}\label{ex 9.1}
Let $n=q+1=3^{3}+1=28=7\times 4,$ $r=4,$ $\delta=4,$ $k=8,$ $b=1$. We have
\begin{eqnarray*}
D&=&\{ \alpha^{i}\mid i=0, \pm1, \pm2, \pm3 \dots, \pm8\}, \\
L_{0}&=&\{\alpha^{i}\mid i \mod\, 7=0\}\\
&=&\{\alpha^{i}\mid i=\bm{0, \pm7,} 14 \},\\
L_{1}&=&\{\alpha^{i}\mid i \mod\, 7=1\}\\
&=&\{\alpha^{i}\mid i=\bm{1, 8,} -13, \bm{-6} \},\\
L_{-1}&=&\{\alpha^{i}\mid i \mod\, 7=-1\}\\
&=&\{\alpha^{i}\mid i=\bm{6,} 13, \bm{-8,} \bm{-1} \}.
\end{eqnarray*}
Then the cyclic code $\cC$ with the defining set of zeros $(L_{1}\cup L_{0}\cup L_{-1})\cup D$ is an optimal $27$-ary cyclic $(4, 4)$-LRC.
\end{xmpl}
\begin{xmpl}\label{ex 9.2}
Let $n=q+1=3^{3}+1=28=7\times 4,$ $r=4,$ $\delta=4,$ $k=8,$ $b=1$. We have
\begin{eqnarray*}
D&=&\{ \alpha^{i}\mid i=\pm6, \pm7, \pm8 \dots, \pm13, 14\}, \\
L_{0}&=&\{\alpha^{i}\mid i \mod\, 7=0\}\\
&=&\{\alpha^{i}\mid i=0, \bm{\pm7, 14} \},\\
L_{1}&=&\{\alpha^{i}\mid i \mod\, 7=1\}\\
&=&\{\alpha^{i}\mid i=1, \bm{8, -13, -6} \},\\
L_{-1}&=&\{\alpha^{i}\mid i \mod\, 7=-1\}\\
&=&\{\alpha^{i}\mid i=\bm{6, 13, -8,} -1 \}.
\end{eqnarray*}
Then the cyclic code $\cC$ with the defining set of zeros $(L_{1}\cup L_{0}\cup L_{-1})\cup D$ is an optimal $27$-ary cyclic $(4, 4)$-LRC.
\end{xmpl}
{\em Remark 13:}
\begin{enumerate}
\item  As mentioned previously, in Construction \ref{construction 6} $\thicksim$ Construction \ref{construction 8}, we can choose the $(r, \delta)$-locality sets as some cosets of the group of $\nu^{\prime}$-th roots of unity, where  $m=i_{0}, \pm i_{1}, \pm i_{2}, \dots, \pm i_{(\delta-2)/2}$ and  $i_{0}=0<i_{1}<\dots <i_{(\delta-2)/2}\le r+\delta-2$ with common odd difference $b$ such that $(b,n)=1$, choose  the  consecutive zeros for $D$ with the uniform difference $b$ in the similar way as Theorem \ref{thm 4.5} $\thicksim $\ref{thm 4.7} in Section \ref{2016 ISIT}, then we can obtain the more general constructions.
\item When $\delta=2$, the general constructions stated above reduce to Theorem \ref{thm 4.5} $\thicksim$ \ref{thm 4.7}  in Section \ref{2016 ISIT}.
\end{enumerate}

Based on the above constructions, it is not difficult to obtain the following generalization.
\begin{theorem}
Let $\alpha \in \ff_{q^2}$ be a primitive $n$-th root of unity,
where $n\mid q+1$ and $q$ is odd. Let $b$ be an positive integer such that  $(b, n) = 1$, and $i_{0}=0<i_{1}<\dots <i_{(\delta-2)/2}\le r+\delta-2$ with uniform difference $b$.
\begin{enumerate}
	\item If $n$ is odd, we can obtain optimal cyclic $(r, \delta)$-LRCs as Construction \ref{construction 2}$\thicksim$ Construction \ref{construction 5} ;
	\item If $n$ is even, we can  obtain optimal cyclic $(r, \delta)$-LRCs as Construction \ref{construction 6}$\thicksim$ Construction \ref{construction 8}.
\end{enumerate}
\end{theorem}
\bigskip
\subsubsection{Optimal cyclic $(r, \delta)$-LRCs with odd $\delta$}
Based on the  \emph{$(\cup_{m}L_{m})\cup D$ construction} and odd $\delta$, we have to choose pairs of  $(r, \delta)$-locality sets $L_{\pm m}$ except $L_0$,  which means $b$ is even. For simplicity, we choose $b=2$  such that  $(b, n) = 1$ in the following constructions.

\begin{cnstr}\label{construction 9}
Let $\alpha \in \ff_{q^2}$ be a primitive $n$-th root of unity,
where $n\mid q+1$ and $q$ is even. If $\mu=k/r$ is  odd, consider the following sets of elements of $\ff_{q^2}$:
\begin{eqnarray*}
D&=&\{\alpha^{\pm 1},  \alpha^{\pm3}, \alpha^{\pm5}, \dots, \alpha^{\pm(1+\frac{n-k-(\mu-1)(\delta-1)-2}{2}\times 2)}\},\\
L_{m}&=&\{\alpha^{i}\mid i \mod\, (r+\delta-1)=m\}, m=\pm 1, \pm 3, \pm 5,  \dots, \pm (\delta-2).
\end{eqnarray*}
Then the cyclic code $\cC$ with the defining set of zeros $(\cup_{m}L_{m})\cup D$ is an optimal $q$-ary cyclic $(r, \delta)$-LRC.
\end{cnstr}

The proof of this construction is similar to Construction \ref{construction 4}.
\begin{xmpl}\label{ex 10}
Let $n=q+1=2^{6}+1=65=5\times 13$, $r=3$, $\delta=3$, $k=21$, $b=2$. We have
\begin{eqnarray*}
D&=&\{ \alpha^{i}\mid i= \pm1, \pm3, \pm5 \dots, \pm31\}, \\
L_{1}&=&\{\alpha^{i}\mid i \mod\, 5=1\}\\
&=&\{\alpha^{i}\mid i=\bm{1,} 6, \bm{11,} 16, \bm{21,} 26, \bm{31,} \bm{-29,} -24, \bm{-19,} -14, \bm{-9,} -4 \},\\
L_{-1}&=&\{\alpha^{i}\mid i \mod\, 5=1\}\\
&=&\{\alpha^{i}\mid i=4, \bm{9,} 14, \bm{19,} 24, \bm{29,} \bm{-31,} -26, \bm{-21,} -16, \bm{-11,} -6, \bm{-1} \}.
\end{eqnarray*}
Then the cyclic code $\cC$ with the defining set of zeros $(L_{1}\cup L_{-1})\cup D$ is an optimal $64$-ary cyclic $(3, 3)$-LRC.
\end{xmpl}

\begin{cnstr}\label{construction 10}
Let $\alpha \in \ff_{q^2}$ be a primitive $n$-th root of unity, where $n\mid q+1$ and $q$ is even. If $\mu=k/r$ is  even, consider the following sets of elements of $\ff_{q^2}$:
\begin{eqnarray*}
D&=&\{\alpha^{0},  \alpha^{\pm 2}, \alpha^{\pm4},\dots, \alpha^{\pm(\frac{n-k-(\mu-1)(\delta-1)-1}{2})\times 2}\},\\
L_{m}&=&\{\alpha^{i}\mid i \mod\, (r+\delta-1)=m\}, m=\pm 1, \pm 3, \pm 5,  \dots, \pm (\delta-2).
\end{eqnarray*}
Then the cyclic code $\cC$ with the defining set of zeros $(\cup_{m}L_{m})\cup D$ is an optimal $q$-ary cyclic $(r, \delta)$-LRC.
\end{cnstr}

The proof of this construction is similar to Construction \ref{construction 5}.

{\em Remark 14:}
  As mentioned previously, in Construction \ref{construction 9} $\thicksim$ Construction \ref{construction 10}, we can choose the $(r, \delta)$-locality sets as some cosets of the group of $\nu^{\prime}$-th roots of unity, where  $m=\pm i_{1}, \pm i_{2}, \dots, \pm i_{(\delta-1)/2}$ and  $0<i_{1}=b/2<i_{2}<\dots <i_{(\delta-1)/2}\le r+\delta-2$ with uniform even difference $b$ such that $(b,n)=1$, choose  the  consecutive zeros for $D$ with the same common difference $b$ in the similar way, then we can obtain the more general constructions.
\begin{xmpl}\label{ex 11}
Let $n=q+1=2^{6}+1=65=5\times 13$, $r=3$, $\delta=3$, $k=24$, $b=2$. We have
\begin{eqnarray*}
D&=&\{ \alpha^{i}\mid i=0, \pm2, \pm4, \pm6 \dots, \pm26\}, \\
L_{1}&=&\{\alpha^{i}\mid i \mod\, 5=1\}\\
&=&\{\alpha^{i}\mid i=1,\bm{ 6,}  11, \bm{16,}  21, \bm{26,} 31, -29, \bm{-24,} -19, \bm{-14,} -9, \bm{-4} \},\\
L_{-1}&=&\{\alpha^{i}\mid i \mod\, 5=1\}\\
&=&\{\alpha^{i}\mid i=\bm{4,} 9, \bm{14,} 19, \bm{24,} 29, -31, \bm{-26,}  -21, \bm{-16,} -11, \bm{-6,} -1 \}.
\end{eqnarray*}
Then the cyclic code $\cC$ with the defining set of zeros $(L_{1}\cup L_{-1})\cup D$ is an optimal $64$-ary cyclic $(3, 3)$-LRC.
\end{xmpl}

Based on the above two constructions, it is not difficult to obtain the following generalization.
\begin{theorem}
Let $\alpha \in \ff_{q^2}$ be a primitive $n$-th root of unity, where $n\mid q+1$ and $q$ is odd. Let $b$ be an positive even integer such that  $(b, n) = 1$, and $0<i_{1}=b/2<i_{2}<\dots <i_{(\delta-1)/2}\le r+\delta-2$ with uniform difference $b$.
If $n$ is odd, we can obtain optimal cyclic $(r, \delta)$-LRCs as Construction \ref{construction 9}$\thicksim$ Construction \ref{construction 10}.
\end{theorem}

\begin{table}[!hbp]\label{table3}
\centering
\caption{\bf Optimal Cyclic $(r, \delta)$ LRCs over $\ff_{q}$ with length $n\mid q+1$ }
\begin{tabular}{|c|c|c|c|}
\hline
 $\delta$&Conditions &Consecutive zeros set $D$ & $(r, \delta)$-locality sets $L_{m}$ \\
\hline
\multirow{3}{*}{} & $q$ even, $\mu$ even; &\multirow{3}{*}{$\lbrace \alpha^{i}| \frac{k+(\mu-1)(\delta-1)+1}{2}\le i\le \frac{2n-k-(\mu-1)(\delta-1)-1}{2}\rbrace$}&\multirow{2}{*}{$m=0, \pm1, \pm2,$}\\
&or $q$ odd,  $\mu$ even, & & \\
&$n$ odd. ($b=1$)& & $\dots, \pm(\delta-2)/2$  \\
 \cline{2-4}
\multirow{3}{*}{} &$q$ even, $\mu$ odd; &\multirow{3}{*}{$\lbrace \alpha^{i}| -\frac{n-k-(\mu-1)(\delta-1)-1}{2}\le i\le\frac{n-k-(\mu-1)(\delta-1)-1}{2}\rbrace$}&\multirow{2}{*}{$m=0, \pm1, \pm2, $} \\
& or $q$ odd, $\mu$ odd, & & \\
& $n$ odd. ($b=1$)&&$\dots, \pm(\delta-2)/2$\\
 \cline{2-4}
\multirow{3}{*}{} &$q$ even, $\mu$ even; &\multirow{3}{*}{$\lbrace \alpha^{2i+1}| -\frac{n-k-(\mu-1)(\delta-1)}{2}\le i \le\frac{n-k-(\mu-1)(\delta-1)-2}{2}\rbrace$} &\multirow{3}{*}{$m=0, \pm2, \pm4,$ }\\
&or $q$ odd, $\mu$ even,& &\\
&$n$ odd. ($b=2$) & &$\dots, \pm(\delta-2)$ \\
 \cline{2-4}
\multirow{3}{*}{even} &$q$ even, $\mu$ odd; &\multirow{3}{*}{$\lbrace \alpha^{2i}| -\frac{n-k-(\mu-1)(\delta-1)-1}{2}\le i\le\frac{n-k-(\mu-1)(\delta-1)-1}{2}\rbrace$}& \multirow{3}{*}{$m=0, \pm2, \pm4, $ }\\
&or $q$ odd, $\mu$ odd,  & &\\
&$n$ odd. ($b=2$)& &$\dots, \pm(\delta-2)$\\
 \cline{2-4}
\multirow{2}{*}{} &$q$ odd, $\mu$ and $\nu^{\prime}$ odd, &\multirow{2}{*}{$\lbrace \alpha^{i}| -\frac{n-k-(\mu-1)(\delta-1)-1}{2}\le i\le\frac{n-k-(\mu-1)(\delta-1)-1}{2}\rbrace$} &\multirow{2}{*}{$m=0, \pm1, \pm2, $}\\
&$n$ even. ($b=1$)& &$\dots, \pm(\delta-2)/2$\\
 \cline{2-4}
\multirow{2}{*}{} &$q$ odd, $\mu$ even, &\multirow{2}{*}{$\lbrace \alpha^{i}|  \frac{k+(\mu-1)(\delta-1)+1}{2}\le i\le \frac{2n-k-(\mu-1)(\delta-1)-1}{2}\rbrace$} &\multirow{2}{*}{$m=0, \pm1, \pm2,$}\\
 &$\nu^{\prime}$ odd, $n$ even. ($b=1$)& &$ \dots, \pm(\delta-2)/2$\\
 \cline{2-4}
&\multirow{2}{*}{} &\multirow{2}{*}{$\lbrace  \alpha^{i}| -\frac{n-k-(\mu-1)(\delta-1)-1}{2}\le i\le\frac{n-k-(\mu-1)(\delta-1)-1}{2}\rbrace$}&\multirow{2}{*}{$m=0, \pm1, \pm2,$}\\
&$q$ odd, $\mu$ and $\nu^{\prime}$ even,  & &\\
 \cline{3-3}
& &\multirow{2}{*}{$\lbrace \alpha^{i}|  \frac{k+(\mu-1)(\delta-1)+1}{2}\le i\le \frac{2n-k-(\mu-1)(\delta-1)-1}{2}\rbrace$}&\multirow{2}{*}{ $\dots, \pm(\delta-2)/2$} \\
 &$n$ even. ($b=1$)  & &\\
 \hline
\multirow{3}{*}{} & $q$ even, $\mu$ odd; &\multirow{3}{*}{$\lbrace \alpha^{2i+1}| -\frac{n-k-(\mu-1)(\delta-1)}{2}\le i \le\frac{n-k-(\mu-1)(\delta-1)-2}{2}\rbrace$}&\multirow{2}{*}{$m=\pm1, \pm3, \pm5, $ } \\
&or $q$ odd,  $\mu$ odd, & & \\
&$n$ odd. ($b=2$)& &$\dots, \pm(\delta-2)$\\
 \cline{2-4}
odd &$q$ even, $\mu$ even; &\multirow{3}{*}{$\lbrace \alpha^{2i}| -\frac{n-k-(\mu-1)(\delta-1)-1}{2}\le i\le\frac{n-k-(\mu-1)(\delta-1)-1}{2}\rbrace$}& \multirow{3}{*}{$m=\pm1, \pm3, \pm5,$ }\\
& or $q$ odd, $\mu$ even, & & \\
& $n$ odd. ($b=2$)&&$ \dots, \pm(\delta-2)$\\
 \hline
\end{tabular}
\end{table}

\section{Conclusion}\label{conclusion}
In this paper, we have studied the constructions of optimal cyclic $(r,\delta)$-LRCs which meet the generalized Singleton-like bound (\ref{eq_GeneralizedSingleton}). Inspired by works of Tamo, Barg, Goparaju and Calderbank \cite{Tamo cyclic},
we firstly generalize their cyclic Reed-Solomon-like $r$-local LRCs to the cases of $(r,\delta)$-LRCs ($\delta\ge 2$), and obtain a class of optimal $q$-ary cyclic $(r,\delta)$-LRCs ($\delta\ge 2$) with lengths $n\mid q-1$. Then, based on the Berlekamp-Justesen codes, we construct a new class of optimal $q$-ary cyclic $r$-local LRCs with lengths $n\mid q+1$ and a new class of optimal $q$-ary cyclic $(r,\delta)$-LRCs ($\delta\ge 2$) with lengths $n\mid q+1$. The proposed optimal cyclic LRCs with lengths $n\mid q+1$ exist only for certain parameters (see Table II and Table III), it is interesting to study whether the constructions  for other parameters exist or not. Related to the famous MDS conjecture \cite{MacWilliams}, one could also consider the following problem: for given $q$, $k$, $r$ and $\delta$, find the largest value of $n$ for which there exist $q$-ary optimal cyclic $(r, \delta)$-LRCs with length $n$. The results of this paper give a step forward.
Future works might also include finding more optimal cyclic LRCs with respect to other bounds, e.g., ones given in \cite{Wang,Itzhak,Cadambe:IT}.


%


%





\begin{thebibliography}{1}

\bibitem{gopalan2011locality}
P.~Gopalan, C.~Huang, H.~Simitci, and S.~Yekhanin, ``On the locality of
  codeword symbols,'' in \emph{IEEE Trans. Inf. Theory}, vol 58, no.~11, pp. 6925--6934,  Nov. 2012.

\bibitem{locality2}
D.~S.~Papailiopoulos, J.~Luo, A.~G.~Dimakis, C.~Huang, and J.~Li, ``Simple regenerating codes: Network coding for cloud storage,'' in
\emph{Proc. IEEE INFOCOM}, Mar. 2012, pp.~2801--2805.


\bibitem{network}
A.~G.~Dimakis, P.~B.~Godfrey, Y.~Wu, M.~J.~Wainwright, and K.~Ramchandram, ``Network coding for distributed storage systems,''  in \emph{IEEE Trans. Inf. Theory}, vol. 50, no. 9, pp. 4539-4551, Sep. 2010.

\bibitem{MDS array codes}
I.~Tamo, Z.~Wang and J.~Bruck, ``MDS array codes with optimal rebuiding,''
in \emph{Proc. Int. Symp. Inf. Theory (ISIT)}, St. Petersburg, Russia. Jul/Aug. 2011, pp.~1240--1244.


\bibitem{Huang pyramid codes}
C.~Huang, M.~Chen, and J.~Li, ``Pyramid codes: Flexible schemes to trade space for access efficiency in reliable data storage systems,'' ¡± in
\emph{Proc. 6th IEEE Int. Symp. NCA}, Feb. 2007, pp.~79--86.

\bibitem{Tamo 2013ISIT}
I.~Tamo, D.~S. Papailiopoulos, and A.~G. Dimakis, ``Optimal locally
repairable codes and connections to matroid theory,''  in \emph{Proc. Int. Symp. Inf. Theory (ISIT)}, Istanbul, Turkey, Jul. 2013, pp.~1814--1818.

\bibitem{Tamo matroidIT}
I.~Tamo, D.~S. Papailiopoulos, and A.~G. Dimakis, ``Optimal locally
repairable codes and connections to matroid theory,'' to be published in \emph{IEEE Trans. Inf. Theory}, 2016.

\bibitem{tam14a}
I.~Tamo and A.~Barg, ``A family of optimal locally recoverable codes,'' in \emph{IEEE Trans. Inf. Theory}, vol. 60, no. 8, pp. 4661--4676, Aug. 2014.

\bibitem{gop14}
S. Goparaju and R. Calderbank, ``Binary cyclic codes that are locally repairable,''  in \emph{Proc. Int. Symp. Inf. Theory (ISIT)}, Honolulu, HI, USA, Jul. 2014, pp. 676--680.

\bibitem{PHuang ITW}
P.~Huang, E.~Yaakobi, H.~Uchikawa, and P.~Siegel, ``Cyclic linear binary locally repairable codes,'' in \emph{Proc. 2015 IEEE Inf. Theory Workshop (ITW),} Jerusalem, Apr. 2015, pp. 1--5.

\bibitem{Tamo cyclic}
I.~Tamo, A.~Barg, S. Goparaju and R. Calderbank, ``Cyclic LRC codes and their subfield subcodes,'' in \emph{Proc. Int. Symp. Inf. Theory (ISIT)}, Hong Kong, China, Jul. 2015, pp. 1262--1266.

\bibitem{Zeh ITW}
A.~Zeh and E.~Yaakobi,  ``Optimal linear cyclic locally repairable codes over small fields,''
in \emph{Proc. 2015 IEEE Inf. Theory Workshop (ITW),} Jerusalem, Apr. 2015, pp. 1--5.

\bibitem{prakash2012optimal}
N.~Prakash, G.~M. Kamath, V.~Lalitha, and P.~V. Kumar, ``Optimal linear
codes with a local-error-correction property,'' in \emph{Proc. Int. Symp. Inf. Theory (ISIT)}, Cambridge, MA, U.S.A., Jul. 2012, pp.~2776--2780.

\bibitem{song14}
W.~Song, S.~Day, C.~Yuen, and T.~Li, ``Optimal locally repairable linear
  codes,'' \emph{IEEE J. Selected Areas Comm.}, vol.~32, pp. 6925--6934, 2014.



\bibitem{papil}
D. S. Papailiopoulos and A. G. Dimakis, ``Locally repairable codes,'' \emph{IEEE Trans. Inf. Theory}, vol. 60, no. 10, pp. 5843--5855, Oct. 2014.

\bibitem{ankit-securelrc}
A. S. Rawat, O.O. Koyluoglu, N. Silberstein,and S. Vishwanath, ``Optimal locally repairable and secure codes for distributed storage systems,'' \emph{IEEE Trans. Inf. Theory}, vol. 60, no. 1, pp. 212--236, Jan. 2014.

\bibitem{kumar IT}
N.~Prakash, G.~M. Kamath, V.~Lalitha, and P.~V. Kumar, ``Codes with local regeneration and erasure correction,'' in \emph{IEEE Trans. Inf. Theory}, vol. 60, no. 8, pp. 4637--4660, Aug. 2014.

\bibitem{Ernvall2016IT}
T. Ernvall, T. Westerbäck,  R. Freij-Hollanti, and C. Hollanti, ``Constructions and properties of linear locally repairable codes,'' \emph{IEEE Trans. Inf. Theory}, vol. 62, no. 3, pp. 1129--1143, Mar. 2016.

\bibitem{Wang}
A. Wang and Z. Zhang, ``Repair locality with multiple erasure tolerance,'' \emph{IEEE Trans. Inf. Theory}, vol. 60, no. 11, pp. 6979--6987, Nov. 2014.

\bibitem{Ankit}
A. S. Rawat and D. S. Papailiopoulos, A.G. Dimakis, and S. Vishwanath, ``Locality and availability in distributed Storage,'' in \emph{Proc. Int. Symp. Inf. Theory (ISIT)}, Honolulu, HI, USA, Jul. 2014, pp. 681--685.

\bibitem{Itzhak}
I. Tamo and A. Barg, ``Bounds on locally recoverable codes with multiple recovering sets,'' in \emph{Proc. Int. Symp. Inf. Theory (ISIT)}, Honolulu, HI, USA, Jul. 2014, pp 691--695.


\bibitem{WangCon}
A. Wang and Z. Zhang, ``Achieving arbitrary locality and availability in binary codes,'' in \emph{Proc. Int. Symp. Inf. Theory (ISIT)}, Hong Kong, China, Jul. 2015, pp 1866--1870.


\bibitem{Cadambe:IT}
V. R. Cadambe and A. Mazumdar, ``Bounds on the size of locally recoverable codes,'' in \emph{IEEE Trans. Inf. Theory}, vol. 61, no. 11, pp. 5787--5794, Nov. 2015.

\bibitem{anticode}
N. Silberstein  and A. Zeh, ``Optimal binary locally repairable codes via anticodes,'' in \emph{Proc. Int. Symp. Inf. Theory (ISIT)}, Hong Kong, China, Jun. 2015, pp. 1247--1251.

\bibitem{some-results}
J. Hao, S.-T. Xia and Bin Chen, ``Some results on optimal locally repairable codes,''  in \emph{Proc. Int. Symp. Inf. Theory (ISIT)}, Barcelona, Spain, Jul. 2016, pp. 440--444.

\bibitem{Berlekamp}
E.~Berlekamp and J.~Justesen, ``Some long cyclic linear binary codes are not so bad,'' in \emph{IEEE Trans. Inf. Theory}, vol.20, no. 3, pp. 351--356, May. 1974.


\bibitem{Roth}
R. M. Roth  and G. Seroussi, ``On cyclic MDS codes of length $q$ over $\ff_{q}$,'' in \emph{IEEE Trans. Inf. Theory}, vol.32, no. 2, pp. 284--285, Mar. 1986.

\bibitem{Georgiades}
J. Georgiades, ``Cyclic $(q+1, k)$ codes of odd order q are not optimal,'' in \emph{Proc. Atti. Sem. Mat. Fis. Univ. Monda},  vol.30, 284--285, 1982.


\bibitem{Cyclic and Pseudo}
C. Dahl and J. P. Pedersen, ``Cyclic and pseudo-cyclic MDS codes of length $q+1$,'' \emph{Journal of Combinatorial Theory,} vol.59, no.1, pp. 130--133, 1992.

\bibitem{MacWilliams}
F. J. MacWilliams and N. J. A. Sloane, ``The Theory of Error-Correcting Codes.''  Amsterdam, The Netherlands: North-Holland, 1981 (3rd printing).

 \bibitem{Roman}
S.~Roman, ``Coding and Information Theory,'' Vol. 134, Springer-Verlag, Berlin, 1992.
\end{thebibliography}
\end{document}